\documentclass[twocolumn]{aastex62}

\graphicspath{{./}{figures/}}

\accepted{July 5, 2022}
\submitjournal{\apj}

\shorttitle{Large Scale Coherence}
\shortauthors{Kim et al.}

\begin{document}

\title{Unexpected dancing partners: Tracing the coherence between the spin and motion of dark matter halos}

\author{Yigon Kim}
\altaffiliation{E-mail: yigon@outlook.com}
\affil{Department of Astronomy and Atmospheric Sciences \\
Kyungpook National University \\
Daegu, Republic of Korea}

\author{Rory Smith}
\altaffiliation{Corresponding author. E-mail: rorysmith274@gmail.com}
\affiliation{Departamento de F$\acute{i}$sica, Universidad T$\acute{e}$cnica Federico Santa Mar$\acute{i}$ae \\
Avenida Vicu$\tilde{n}$a Mackenna 3939, San Joaqu$\acute{i}$n, Santiago, Chile}

\author{Jihye Shin}
\affiliation{Korea Astronomy and Space Science Institute \\
Daejeon, Republic of Korea}



\begin{abstract}

A recent study conducted using CALIFA survey data \citep{2019ApJ...884..104L} has found that the orbital motions of neighbor galaxies are coherent with the spin direction of a target galaxy on scales of many Megaparsecs. We study this so called `large-scale coherence' phenomena using N-body cosmological simulations. We confirm a strong coherence signal within 1~Mpc/h of a target galaxy, reaching out to 6~Mpc/h. We divide the simulation halos into subsamples based on mass, spin, merger history and local halo number density for both target and neighbor halos. We find a clear dependency on the mass of the target halo only. Another key parameter was the local number density of both target and neighbor halos, with high density regions such as clusters and groups providing the strongest coherence signals, rather than filaments or lower density regions. However we do not find a clear dependency on halo spin or time since last major merger. The most striking result we find is that the signal can be detected up to~15Mpc/h from massive halos. These result provide valuable lessons for how observational studies could more clearly detect coherence, and we discuss the implications of our results for the origins of large-scale coherence.

\end{abstract}



\section{Introduction} \label{sec:intro}

Thanks to the availability of many large area surveys of galaxies, observed at a wide variety of wavelengths and using various telescopes and instruments such as CfA redshift survey, SDSS, 2dF, 2MASS \citep{1982ApJ...253..423D, 1999PASP..111..438F, 2000AJ....120.1579Y, 2001MNRAS.328.1039C, 2009MNRAS.399..683J, 2012ApJS..199...26H}, our knowledge of the cosmic large scale structure has leapt forward in recent decades. It is now well known that there are a variety of structures in our Universe on scales of several Megaparsecs including filaments, walls, clusters or voids \citep{1986ApJ...306..341G, 2005ApJ...624..463G}. In the concordance model of modern cosmology, these structures are grown from initial fluctuations in the early stages of the Universe \citep{1970A&A.....5...84Z, 1982PhLB..115..295H, 1982PhLB..117..175S, 1983PhRvD..28..679B, 1985ApJ...292..371D, 2004ApJ...601....1W}. Gravity acts to shape the large scale structure, causing collapse in overdense regions (future clusters), and expansion in underdense regions (future voids). This process generates the web-like structures such as the dense knots, filaments, walls and the nearly empty voids that we observe today.

While this evolution of the large-scale structure occurs, individual dark matter halos which are embedded in the structure begin to achieve their own internal spin. Tidal Torque Theory (TTT) is the traditional explanation for the origins of their spins \citep{1951pca..conf..195H, 1969ApJ...155..393P, 1970Ap......6..320D, 1984ApJ...286...38W}. In TTT, protohalos acquire their angular momentum through interactions with nearby massive structures. TTT is a good example of how the large scale structure (LSS) affects individual halos. It is now well known that the halo's location in the LSS affects its spin \citep{2004MNRAS.352..376A, 2005ApJ...627..647B, 2007MNRAS.381...41H, 2008MNRAS.389.1127P, 2012MNRAS.421L.137L, 2013ApJ...762...72T, 2014MNRAS.440L..46A}. Similarly, \citet{2018MNRAS.481.4753C} finds a relation between the halo spin orientation and the direction of filaments in which they are embedded. They claim that this could be the result of a combination of TTT at early times combined with mergers especially at more recent times (see also \citet{2015MNRAS.452.3369C}). Not only is the LSS thought to be responsible for the spin of halos, but there are also some works that claim an environmental density dependency on direction of a galaxy's spin, especially  in early types galaxies \citep{2006MNRAS.366.1126C, 2007MNRAS.379..401E, 2011MNRAS.413..813C, 2011MNRAS.414..888E, 2011MNRAS.414.2923K}. \citet{2015MNRAS.446.2744L} reveals that halos which are embedded in cosmic filaments tend to spin with the vorticity of their host filament. They find that spin alignment with vorticity is present at all halos masses but it is stronger for more massive halos. Similarly, \citet{2020arXiv200506479A} report the presence of an alignment between the spin vectors of fast rotating DM halos and the orbital angular momentum vector of their neighbors (mergers and fly-bys) due to the flows of matter within the LSS, especially within walls and filaments. Summarising, it is now clear that halo spin strength and direction are closely connected with the LSS. But it is less clear under what conditions the coherence between a halo's spin direction and the bulk motion of its neighbouring halos within LSS would be strongest, and how far it might extend.

With the recent advent of observational surveys using integrated field units (IFU) \citep{2012A&A...538A...8S, 2015ApJ...798....7B, 2015MNRAS.447.2857B}, it has been possible to collect individual spectra from spaxels across the disks of galaxies, and a great deal has been learnt about galaxies and their evolution. New insights have been provided into several fields including the internal dynamics (stellar and ionised gas) of individual galaxies, chemical compositions and stellar populations throughout their disks \citep{2015MNRAS.449..867B, 2016MNRAS.463.2799I, 2017MNRAS.465..688G}. With a measurement of the redshift at each location across a galaxy disk, we can construct full kinematic maps of their stellar disks. Using IFU data, a signal of coherence between the spin of the stellar disk of target galaxies and their neighbors' motion has recently been reported (\citet{2019ApJ...872...78L, 2019ApJ...884..104L}; hereafter denoted as L19a and L19b respectively). These two papers report a strong signal of coherence within 800~kpc from target galaxies, but also a weaker signal of coherence stretching out to several megaparsecs. Given its large extent, this rules out that the coherence could be generated by some kind of earlier interaction between the target and its neighbors. Some marginal dependencies were reported on the observed amplitude of the coherence signal. The coherence signal amplitude was greater around galaxies whose optical disks are less concentrated, and those whose inner and outer disk spin direction was more similar. Also, galaxy pairs embedded in dense regions show a larger amplitude of coherence. But, the statistical significance of these dependencies was generally under one-sigma.

In this paper, we conduct a detailed parameter study of the amplitude of coherence between a target halo's spin and their neighbors' motions using N-body cosmological simulations. Our aim is to better understand what dictates the amplitude of coherence, both on short and long distance scales, in the hope of gaining deeper insights into its physical origin, in addition to providing useful feedback for future observations of the phenomena. The structure of this paper can be summarised in the following. In section \ref{sec:data}, we describe our simulations and data. The method for measuring the coherence signal is described in section \ref{sec:method}. In section \ref{sec:result}, the results of measuring coherence including dependencies on the halo mass, spin parameter and environment are considered. In section \ref{sec:discussion} we put our results into context, comparing with the observations, and discussing possible origin scenarios. Finally, in section \ref{sec:conclusion} we summarize and conclude our study.

\section{Data}\label{sec:data}
  
\subsection{Cosmological simulation}

We measure the coherence signal in a set of N-body cosmological simulation, that were run as dark matter only models using the GADGET-3 code \citep{2001NewA....6...79S}. To generate the initial conditions at redshift=200, the Multi Scale Initial Condition software (MUSIC; \citet{2011MNRAS.415.2101H}) was used. The package {\sc{Code for Anisotropies in the Microwave Background}}(CAMB; \citet{2000ApJ...538..473L}) calculated the linear power spectrum. A set of five cubic cosmological volumes were analysed, with dimensions $120\times120\times120$~Mpc/h for each box. The dark matter particles have a fixed mass of $1.072\times10^9 M_{\odot}/h$ in all the simulations. A cosmology of $\Omega_m = 0.3$, $\Omega_{\Lambda} = 0.7$, $\Omega_b = 0.047$ and $h = 0.684$ was assumed for the initial conditions and running of the simulations. Output files were produced at $\sim$100~Myr intervals consisting of the mass, velocities and positions for each snapshot, in comoving coordinates down to redshift zero. However, this study is conducted entirely at redshift z$ = 0$. Where comparison with observations was necessary, we used today's value of Hubble's constant to convert comoving quantities into physical quantities (e.g., distance, mass). The five cosmological boxes were sufficient to provide excellent statistics for our study ($\sim$1.5 million halos). Each box is a different random realization. Only a small variation in the amplitude of the coherence signal was found between the boxes (see section \ref{sec:allsignal}).

\subsection{ROCKSTAR halo finder}\label{sec:rockstar}

We use the ROCKSTAR halo finder to build a catalogue of halos at $z=0$ \citep{2013ApJ...762..109B}. Halos are identified using a hierarchical Friends-of-Friends (FoF) approach that combines six-dimensional phase space information and one-dimensional time, and provides information on the merging history as well. Combining all 5 simulation boxes provides us with excellent statistics, in terms of numbers of halos, for our study. We remove all halos with a mass lower than $10^{11} M_{\odot}$ in order to ensure our halos are sufficiently resolved more than 65 particles per halo. This leaves us with $298,694$ halos (21 per cent of the total halo catalog). ROCKSTAR provides a number of measurements of the halos that we use in this study. First, the mass of each DM halo is calculated based on a spherical overdensity calculation for all matter above 200 times the critical density of the Universe at that time. The spin of each halo is calculated using $\lambda_{B} \equiv {J}/{\sqrt{2GMR}}$, where J is the angular momentum of the halo, M and R corresponds to virial mass and virial radius of halo \citep{2001ApJ...555..240B}. The merger history is also parameterised as the scale factor when the halo experienced its most recent major merger (defined as being more major than a 1:3 merger).

For estimating the skeleton of the filaments we use the DisPerSE code \citep{2011MNRAS.414..350S}, applied to our halo catalogues, with a persistence ratio threshold of 6.2, and smoothing the filaments 10 times using the provided software. The skeletons trace out the backbone of the filaments, and we define our `filament halos' as any halo located within 2~Mpc from the skeleton consistent with the choice of parameters and recent results of \citealp{2022arXiv220109540J}. This choice also provides us with good number statistics when we make a `filament halo' subsample in Section \ref{sec:dependency}.

\section{Methodology}\label{sec:method}

\begin{figure}
    \centering
    \includegraphics[width=\columnwidth]{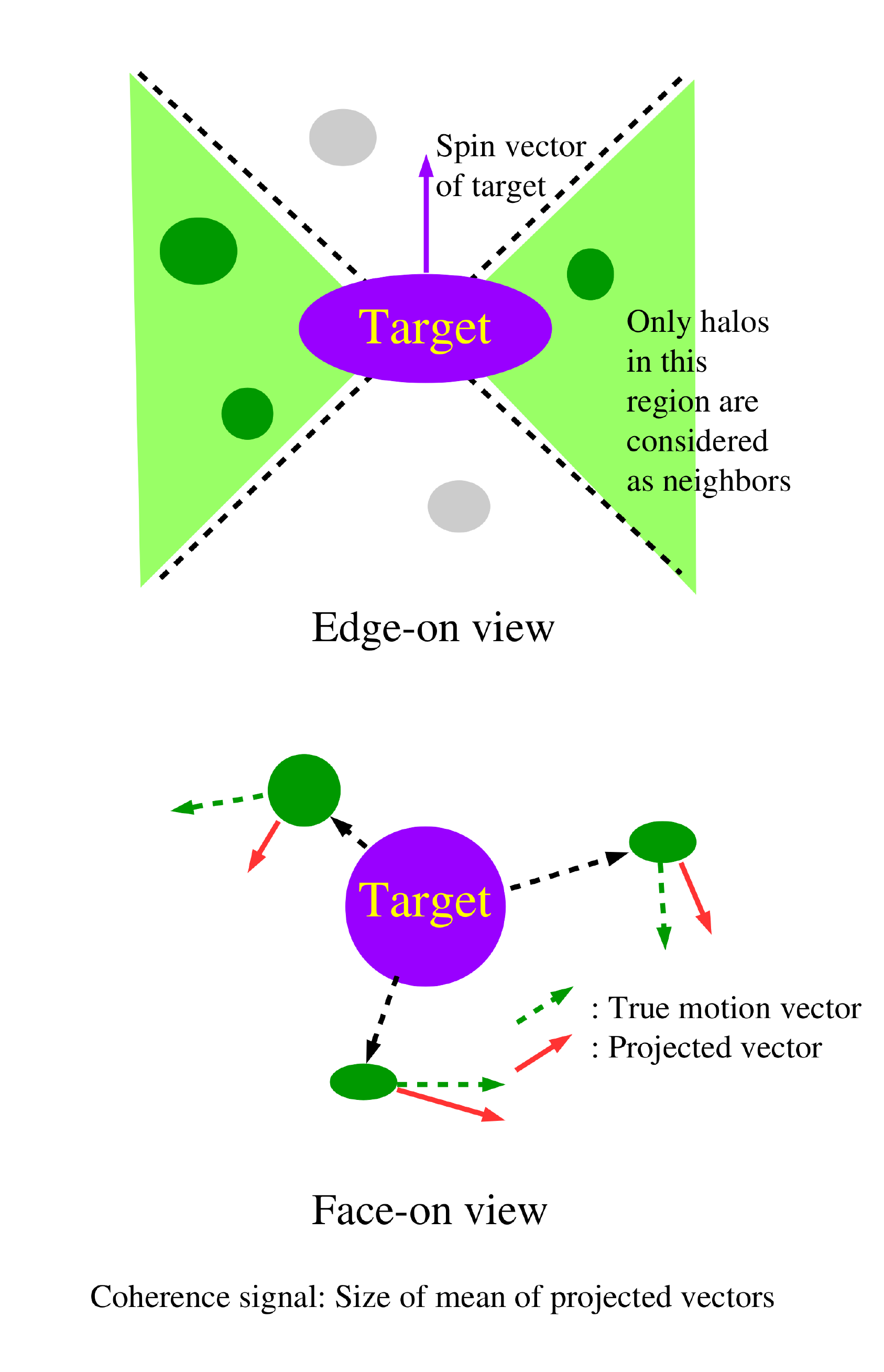}
    \caption{Cartoon schematic of the method used to measure the coherence signal in this study. We only consider neighbors that fall in the green filled region between the dashed lines (where the angle between spin vector and the vector to the neighbor is $>\pi/4$ and $<{3\pi}/4$).}
    \label{fig:cartoon}
\end{figure}

To measure the amplitude of the coherence in the simulation, we used a similar approach as in L19a,b. However, one key difference is that we could use the full 3D information of each halo's movement, while in the observational study of L19a,b only 2D projected information is available. Projection effects will tend to weaken the amplitude of the signal, thus by using the 3D information we are more sensitive to potentially weak coherence signals at large distances. We test the impact of projection effects in section \ref{sec:compare}.

Our method is as follows. First, we select target halos. These are halos with masses between $10^{11}$ and $10^{14} M_{\odot}$. Next, we measure the target halo's spin vector, and rotate all the halos (target and neighbors) so as the spin vector is aligned along the $z$-axis. Then we adopted the X-cut suggested by L19a, that excludes neighbors whose angle between their position vector with respect to the target halo and the target's spin vector is under 45 degrees and over 135 degrees (see Fig. \ref{fig:cartoon}). Since we are using a simulation box rather than real observational data, it is not essential that we use the X-cut but we chose to include it anyhow for better comparison with the results of the observational study. We tested the impact of using the X-cut and find that the signal typically increases in strength by 25 percent, therefore there are benefits to applying this cut. In the final step, we calculate the coherence signal amplitude. For each neighbor, we calculate the tangent velocity with respect to target's spin vector. We then average the tangent velocity over all neighbors within the specified distance range, using a mass weighting. This is analogous to the luminosity-weighting that was used in L19a. Hereafter, we refer to this averaged tangential velocity as the `amplitude' of the coherence. We also refer to the `significance' of the coherence which is separate from the amplitude in that it is a measure of the signal to noise ratio of the coherence signal. The noise is the standard deviation of 100 amplitudes calculated by constructing 100 randomised catalogs for each simulation box. The physical quantities of each halo (such as mass, velocity, size of spin vector and merger tree information) is identical to the original halo catalog, but only the direction of the target halo spin vectors are randomised. In this way, we can determine the significance of the coherence signal we measure. We then split up our sample into different subsamples by various parameters (halo mass, spin, etc) in order to test how the coherence signal depends on these parameters. 

\section{Result}\label{sec:result}

\subsection{The 2D cell plot}\label{sec:cellplot}

\begin{figure*}
    \centering
    \includegraphics[width=130mm]{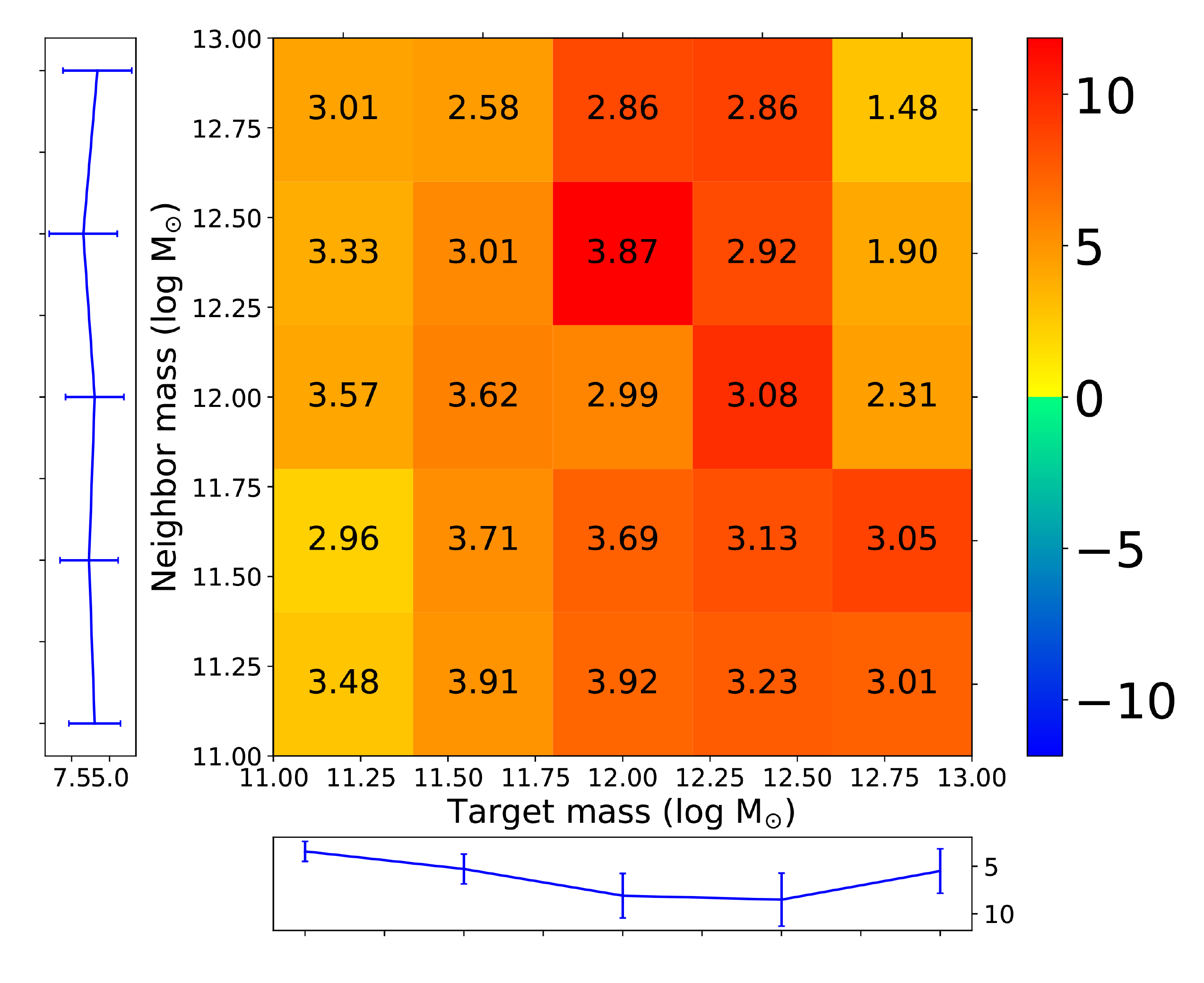}
    \caption{Sample of 2-dimensional cell plot for this study. Coherence amplitude in this plot is measured and integrated within 0 to 6Mpc/h range. The amplitude is the tangential velocity of neighbor halos with respect to target halo, so the units are km/s. The color mean the amplitude of signal at each conditions. In this plot, the condition is DM halo mass. We divided the color scheme for plus and minus signal for easy identification of the positive or negative values. We put the signal to noise ratio for each cell and if the significance is lower than 1, we did not paint the cell. Blue solid lines and error bars show the integrated amplitude along each axis and size of error is obtained from standard deviation of the amplitude from 100 catalogs in which the spin vector of the halos are randomly oriented.}
    \label{fig:sampleplot}
\end{figure*}

This plot was developed to study how the amplitude of the target's coherence signal depends on both target and neighbor halo properties. An example is shown in Fig. \ref{fig:sampleplot}, where the x-axis is the target halo mass and the y-axis is the neighbor halo mass. The color of each cell indicates the amplitude of the coherence signal in each cell, as shown in the color bar. In this plot, the amplitude is measured with all target halos located between 0 to 6~Mpc/h. The small number provided in each cell shows the significance. If the significance is less than 1, the cell is left blank to reduce visible noise. We note that the significance is highly dependent on the number statistics of halos in each of the cells, e.g. we can expect high significance for lower mass halos and lower significance for higher mass halos simply because there are many more low mass halos than high mass halos. The blue solid lines show the integrated coherence signal along each axis, so that we can see the dependence of the signal amplitude on just the quantity on that axis alone, marginalising over the other axis. This histogram helps us to see the dependency on single variable. We do not see a clear dependence on neighbor halo mass (along the y-axis). But, there is a peak in the signal amplitude for target halos of intermediate mass (along the x-axis) in Fig. \ref{fig:sampleplot}

\subsection{Dependency of Signal Amplitude on Distance}
\label{sec:allsignal}

\begin{figure}
    \centering
    \includegraphics[width=\columnwidth]{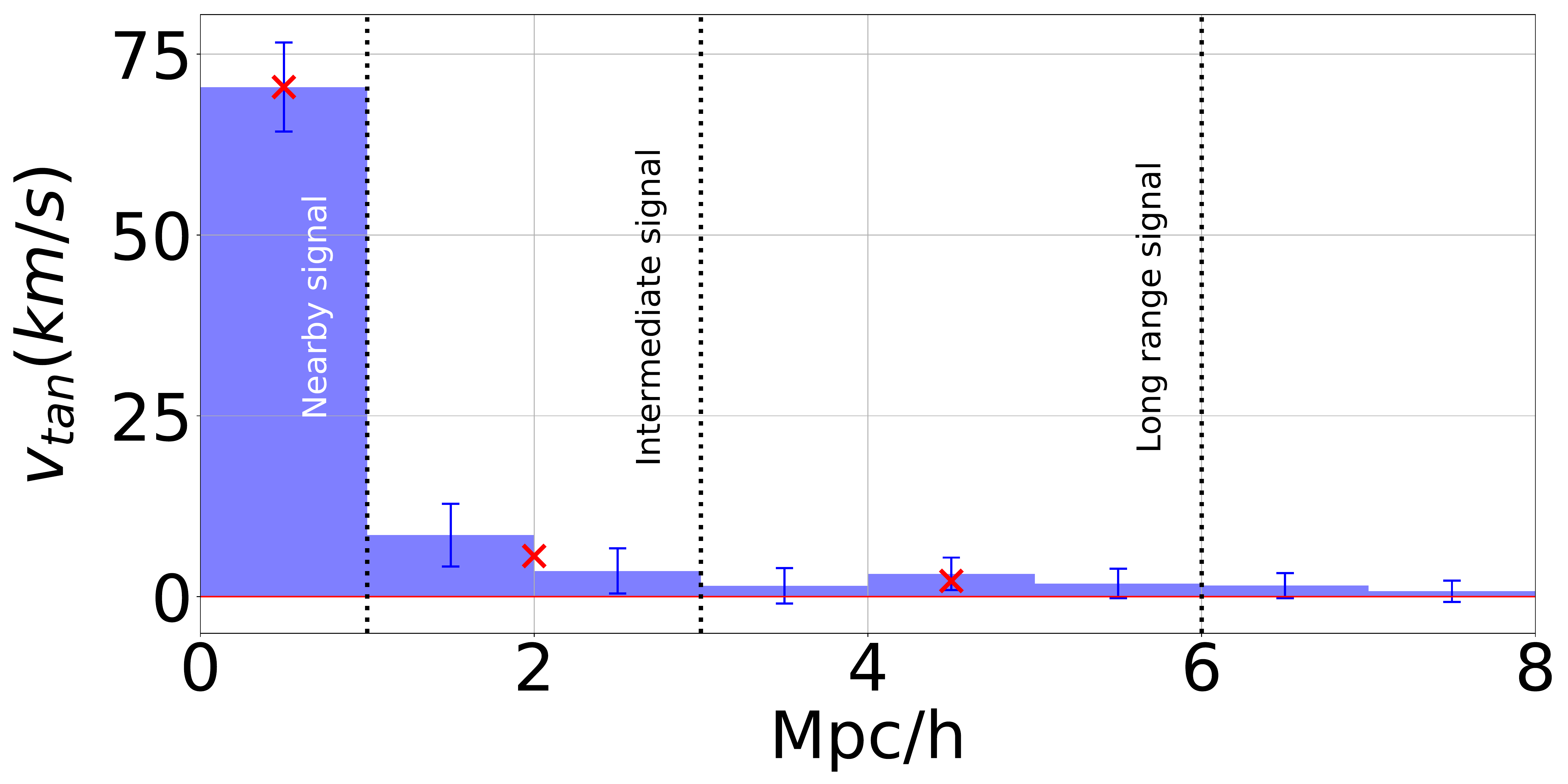}
    \caption{1-Dimensional histogram of the amplitude of the coherence signal of all the target halos in all five simulation boxes. The blue histogram corresponds to the amplitude of the coherence signal in 1~Mpc/h bins, red cross symbols corresponds to integrated amplitude within nearby (0-1~Mpc/h), intermediate (1-3~Mpc/h), and long range (3-6~Mpc/h) distance ranges. The three vertical dotted lines indicate the nearby, intermediate and long range zones. Error bars show the standard deviation of amplitude measured from 100 randomly spin-oriented catalogs to show the signal to noise ratio of the coherence signal. The red horizontal line indicates the zero point.}
    \label{fig:radial}
\end{figure}

\begin{table*}[t]
  \centering
  \begin{tabular}{|c c c c c c c c c|} 
   \hline
    & 0-1Mpc/h & 1-2Mpc/h & 2-3Mpc/h & 3-4Mpc/h & 4-5Mpc/h & 5-6Mpc/h & 6-7Mpc/h & 7-8Mpc/h \\ [0.5ex] 
   \hline
   Amplitude(km/s) & 70.4 & 8.51 & 3.56 & 1.50 & 3.16 & 1.83 & 1.53 & 0.755 \\ 
   \hline
   Significance($\sigma$) & 11.5 & 1.96 & 1.14 & 0.609 & 1.41 & 0.901 & 0.887 & 0.510 \\
   \hline
  \end{tabular}
  \caption{Numerical values of amplitude and significance of the coherence signal shown in \ref{fig:radial}.}
    \label{tab:fig3table}
\end{table*}

In this section, we consider all of our target halos combined and study how the coherence signal amplitude depends on the distance to the neighbor halos, as shown in Fig. \ref{fig:radial}. The amplitude of the coherence is large in the 0 to 1~Mpc/h range, which we refer to as the `nearby signal'. This was also reported observationally in L19a,b. A weak but still statistically significant(S/N $\sim$ 1.5) coherence signal is detected in the 1-3~Mpc/h range. This is slightly larger than the typical virial radius of galaxy clusters \citep{1995ApJ...438..527G, 1997ApJ...478..462C}, meaning that target halos inside clusters remain significantly coherent with neighbors beyond the cluster. We refer to this as `intermediate range coherence'. We refer to the 3-6Mpc/h range as `long range coherence'. Beyond 6Mpc/h, the significance of coherence for all the target halos combined becomes lower than 1. We note that the maximum range of the coherence could be larger than this for a different subsample of target halos.

\begin{figure}
    \centering
    \includegraphics[width=\columnwidth]{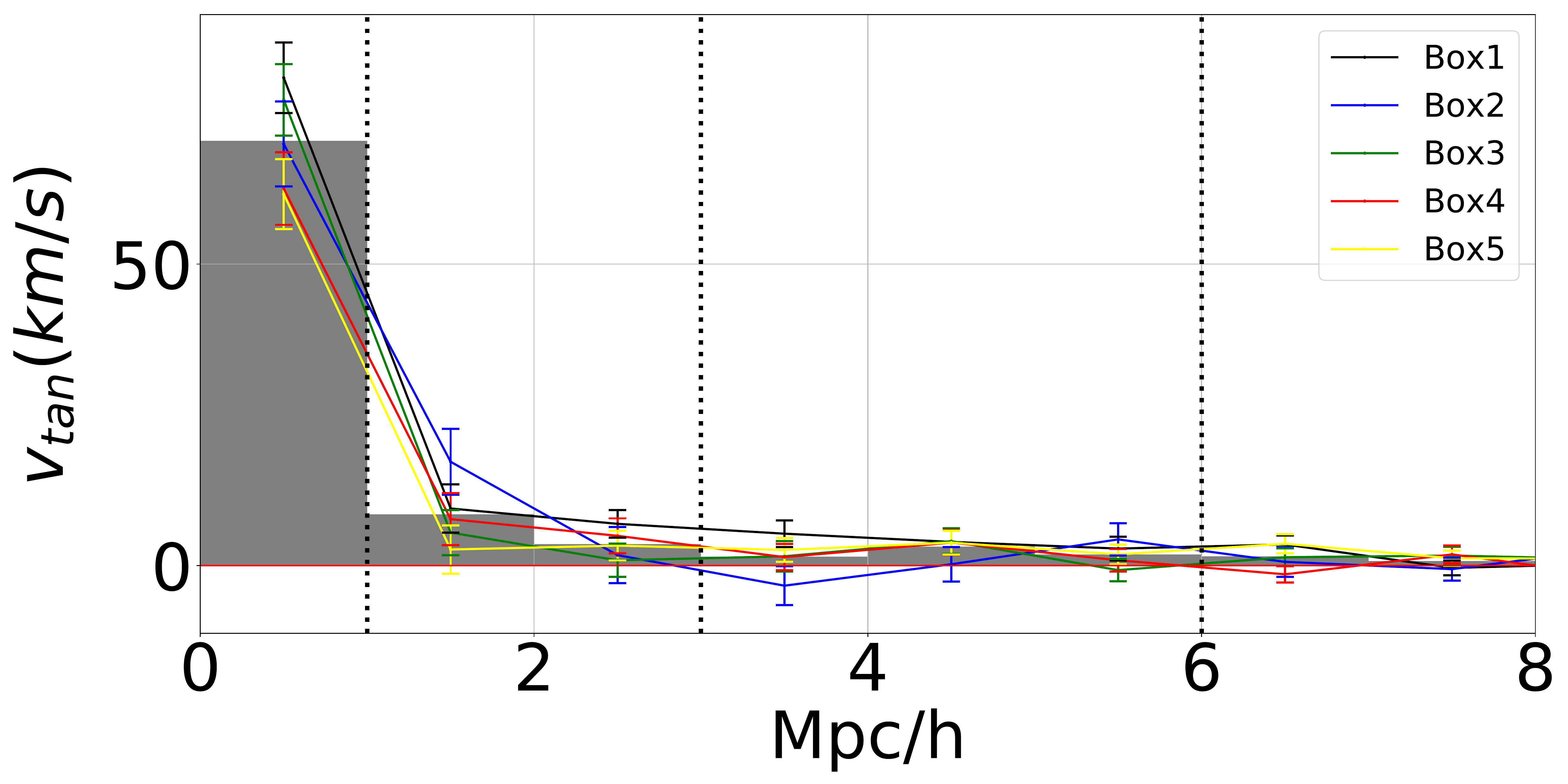}
    \caption{Same format as Fig. \ref{fig:radial}, but here the coherence signal amplitude of each of the five cosmological simulation boxes is plotted individually as a different colored solid lines (see legend). The standard deviation of the amplitude from the 100 catalogues where the spin direction is randomised are plotted as colored error bars. The coherence amplitude of all the boxes combined is plotted as a gray histogram. }
    \label{fig:radial2}
\end{figure}

We also measured the difference in coherence amplitude with distance between each of our individual five cosmological boxes and the result is shown in Fig. \ref{fig:radial2}. The difference is calculated by the ratio between maximum and minimum coherence signal amplitude for each box. The maximum difference in the amplitude was 15 per cent in the 0 to 1~Mpc/h range, 102 per cent in the 1 to 3Mpc/h range, and 167 per cent in the 3 to 6Mpc/h range. Error bars were calculated by creating random-spin catalogs as described in section \ref{sec:method}. The box-to-box differences are generally smaller than one-sigma.

\subsection{Dependency of coherence signal amplitude on halo properties}\label{sec:dependency}
    
\begin{figure*}
    \centering
    \includegraphics[width=150mm]{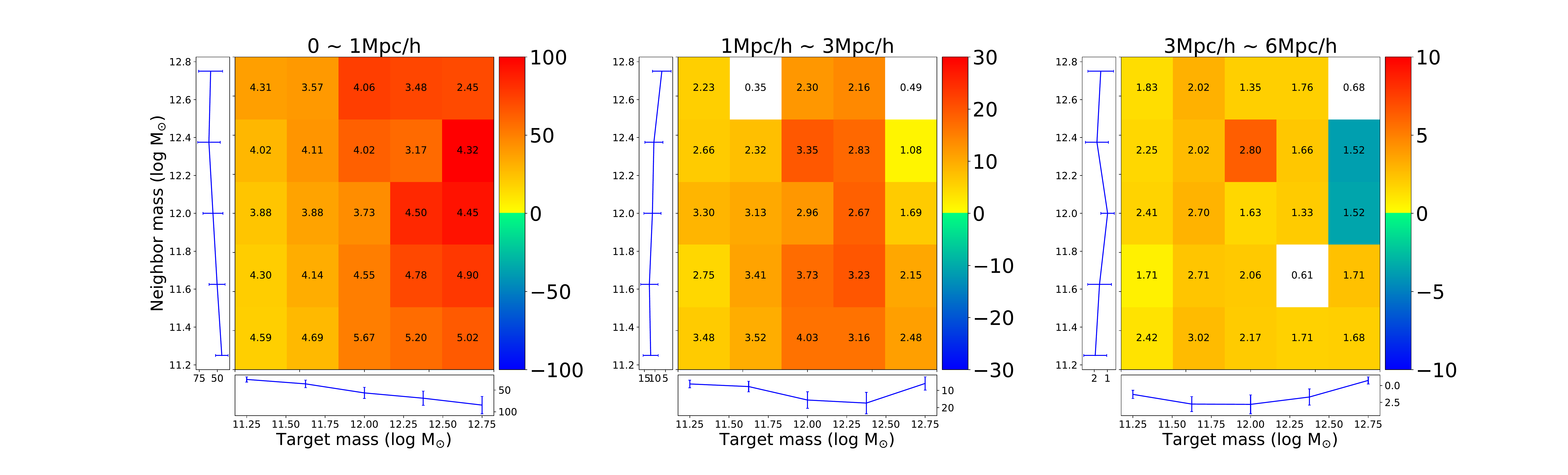}
    \includegraphics[width=150mm]{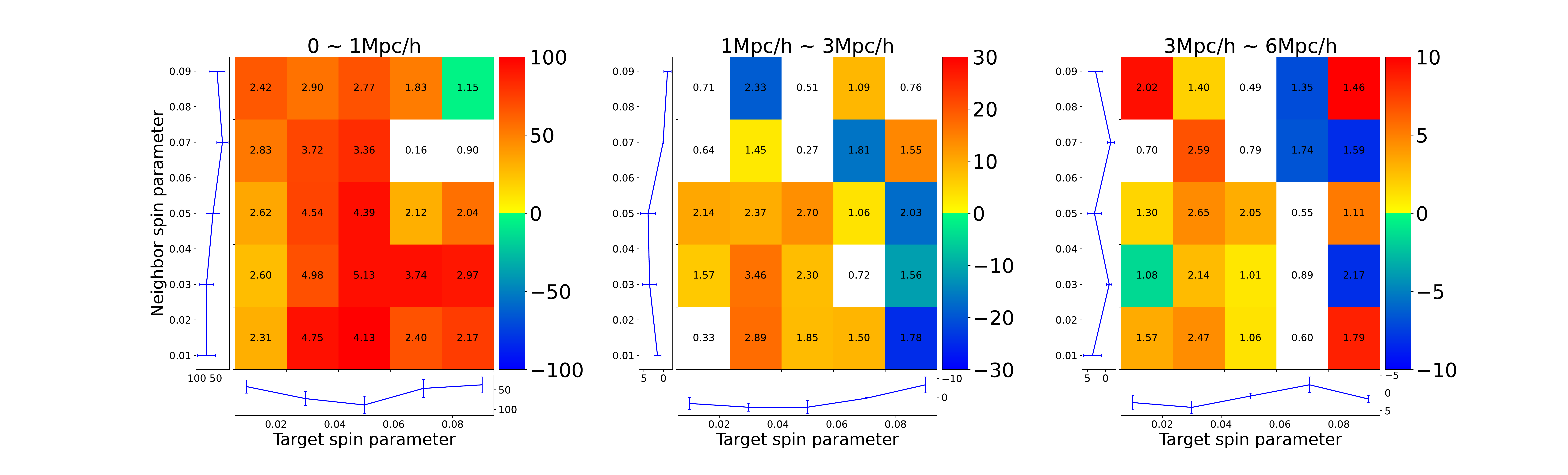}
    \includegraphics[width=150mm]{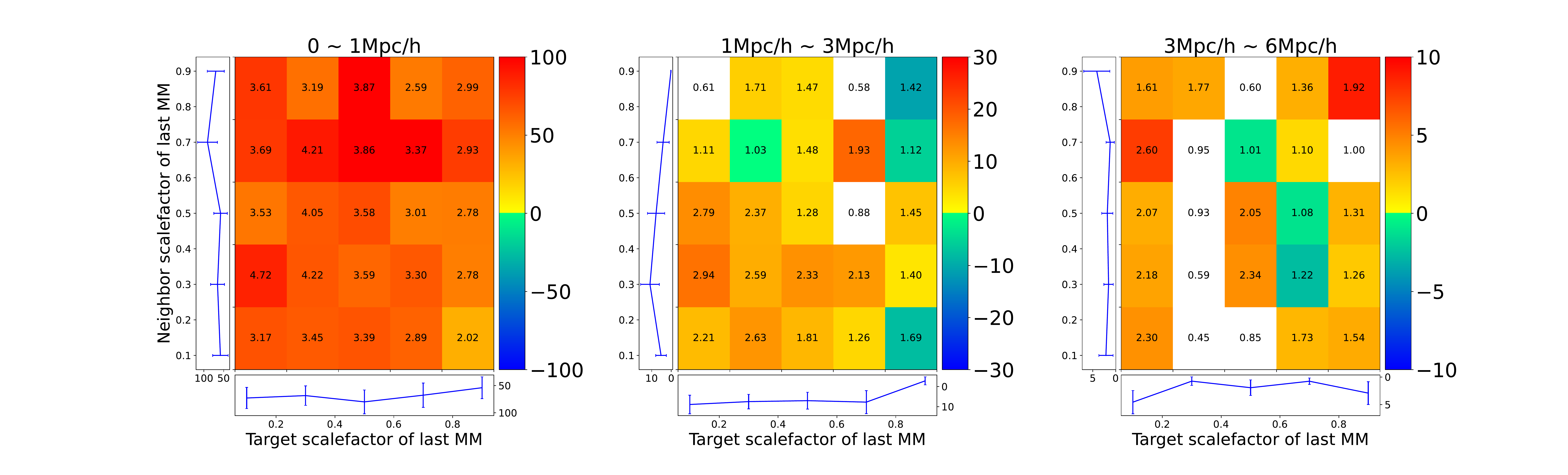}
    \includegraphics[width=150mm]{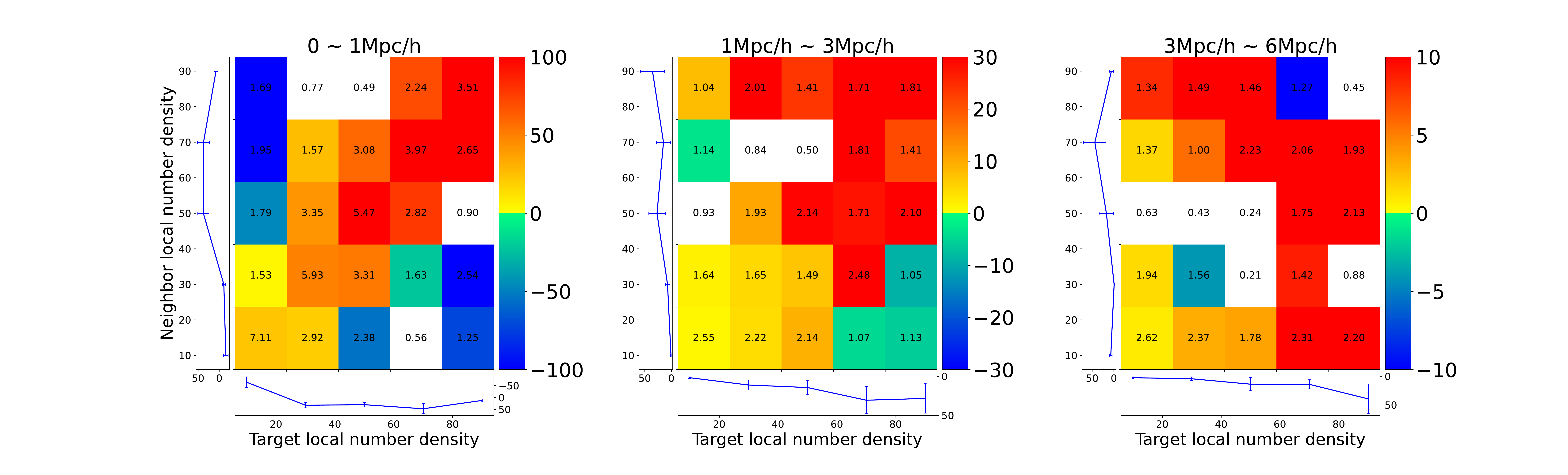}
    \caption{Two dimensional cell plot for visualize how each parameters affect to the coherence. The format is the same as in Fig. \ref{fig:sampleplot} but we selected target halos by four halo properties (top to bottom row; halo mass, halo spin, scale factor of last major merger, and local number density) and the distance range is also divided into three groups in this figures (left to right columns; see panel subtitles).}
    \label{fig:paramdep}
\end{figure*}

Having confirmed the presence of coherence in our simulation boxes for the total sample of target halos, we now divide the halos into subsamples based on various properties of the halos described below. There are four rows in Fig. \ref{fig:paramdep}, one for each halo properties. The columns indicate the different distance ranges we consider between target and neighbor halos (see panel subtitles).
    
\noindent
{\it{Halo mass:}}
We find that the halo mass of the target halos is an important parameter for the amplitude of the coherence signal (see the first row of Fig. \ref{fig:paramdep}). We already saw this dependency on the target mass in Fig. \ref{fig:sampleplot}, but here we can see how this dependency changes as the distance range changes by comparing between the columns. There is a clear and significant dependency on the target halo mass in all the distance ranges, but the dependency on neighbor halo mass is weak. The target mass where the coherence amplitude peaks is different depending on the distance to the neighbors. Within 1~Mpc/h, the amplitude grows linearly up to $10^{13}$ $M_{\odot}$. However with increasing range, the peak target halo mass falls from $10^{12.5}$ to $10^{12.0}$ $M_{\odot}$ (1-3Mpc/h to 3-6Mpc/h, respectively). This mass range corresponds to a stellar mass of $\sim 10^{10}$ $M_{\odot}$ as estimated from halo abundance matching \citep{2010MNRAS.404.1111G}. Thus, our simulations predict that galaxies of this mass would be good target for detecting a high significance and high amplitude signal of coherence.

\noindent
{\it{Halo spin and Scale Factor of Last Major Merger:}} 
In the second row we consider how the amplitude of coherence depends on the spin parameter \citep{2001ApJ...555..240B} of target and neighbor halos. In the third row, we consider how the Scale Factor of the last major merger of the target and neighbor halos affects the coherence amplitude, where a major merger is defined as having a mass ratio of 1:3 or more major. Unlike with the target mass, we do not see any clear dependencies on halo spin and time after last major merging. We will compare this result with the observations in section \ref{sec:compare}.

\noindent
{\it{The local halo number density:}}
In the fourth row of Fig. \ref{fig:paramdep}, we consider how coherence amplitude depends on the local number density of halos. This is defined by counting the total numbers of halos within a 1~Mpc/h sphere around the target or neighbor halos. Thus it is a proxy for the environment in which the halo is embedded. To give a more physical interpretation of these number densities, we test and find that typically the number densities are $\sim$10 around large filaments (measured out to a maximum of 2~Mpc/h from the filament skeleton). Within groups and clusters, there is a large scatter and the number density depends on hostcentric radius too. But typically, the local number density within the virial radius of a host was $\sim$20 for low mass groups (mass $\sim 10^{13.2} M_{\odot}$), $\sim$40 for intermediate mass clusters (mass $\sim 10^{14.2} M_{\odot}$), and $\sim$80 for the most massive clusters (mass $\sim 10^{14.8} M_{\odot}$). 
Although, we do not always see a strong dependency on target or neighbor halo density individually (see 0-1~Mpc/h panel), we clearly see a strong dependency when we consider both the target and neighbor halo number densities together. The coherence amplitude is much higher when both targets and neighbors are located in dense regions, such as if they were both located in the dense cores of clusters. This can clearly be seen by the diagonal stripe (from bottom-left to top-right) visible in the bottom-left panel of Fig. \ref{fig:paramdep}. However, we note that the narrowness of this stripe is partly by construction. If we only measure the coherence amplitude over a 1~Mpc/h range from the target, then it is likely that both the target and neighbor inhabit a similar environmental density. This is why the width of stripe is widened when we consider larger distances of neighbor halos from the target halos (e.g., 1-3 Mpc/h or 3-6 Mpc/h). We see that the signal-to-noise reduces when the cells are located further from the diagonal due to the reduced statistics in each cells. This is especially the case when the target and neighbour halos are close, but less so at larger distances due to the improved statistics. Meanwhile, for close separations between the target and neighbour halo, we see the presence of negative coherence for cell at larger distances from the diagonal strip, as discussed in Section \ref{sec:negacor}.

\begin{figure}
    \centering
    \includegraphics[width=\columnwidth]{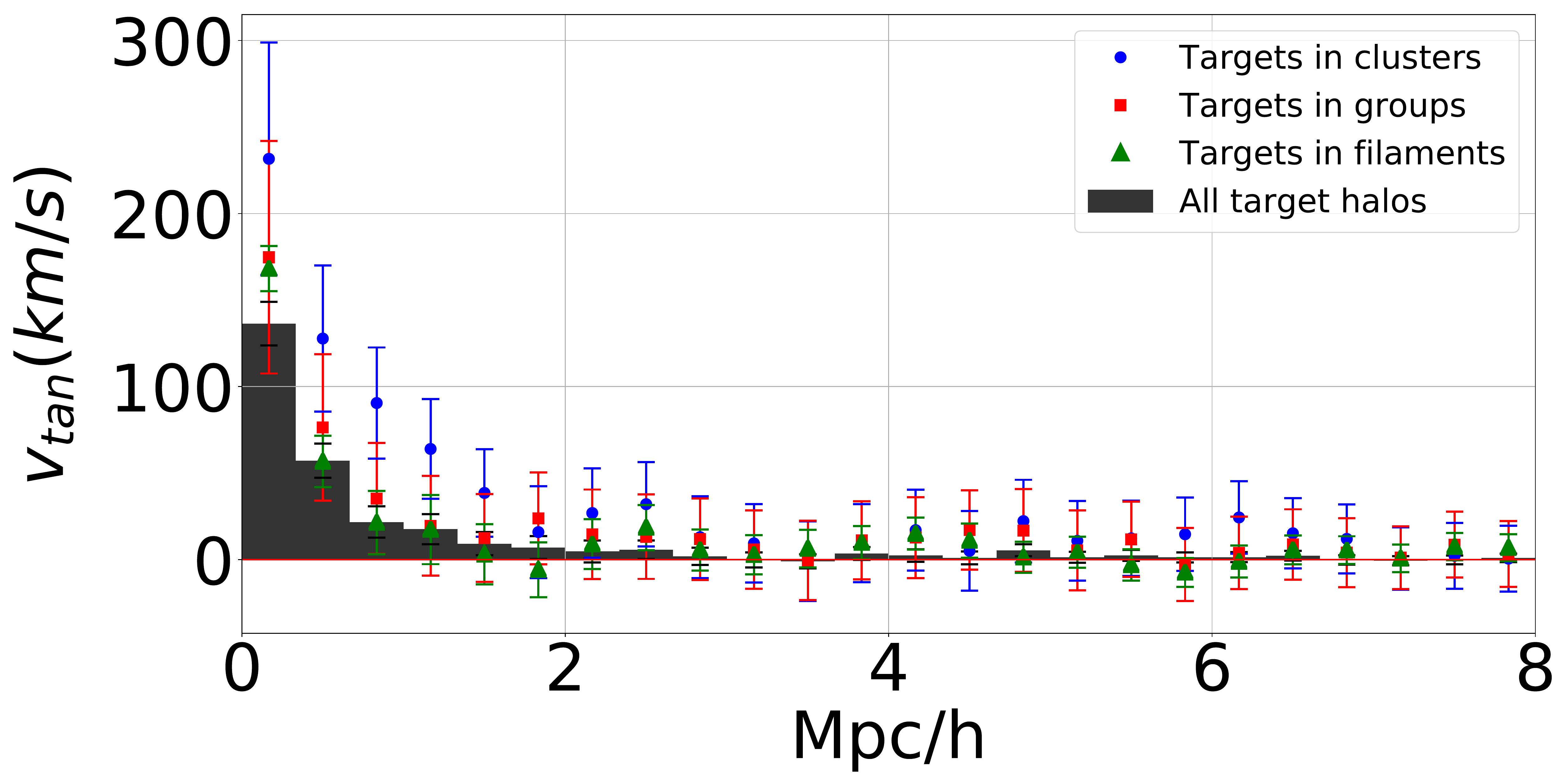}
    \caption{Differences between the amplitudes of the coherence signal in various environment; clusters, groups and filaments. The amplitude of the coherence of all target halos combined is plotted as grey histogram. Blue, red and green markers show the amplitude within clusters, groups and filaments, respectively. The error bars corresponds to the standard deviation of the amplitude measured from the 100 catalogs in which the halo spin direction is randomised.}
    \label{fig:subsmpl_env}
\end{figure}

Since we have seen that the environmental density is an important parameter, we now select targets in specific environments, including clusters, groups and filaments. Clusters are identified as halos with mass $>10^{14} M_{\odot}$, groups have masses between $10^{13}$ and $10^{14} M_{\odot}$. Targets that are considered to be in clusters (groups) are selected out to twice the virial radius of the cluster (group), in order to improve statistics and account for backsplash galaxies. For filament halos, we select halos that are located within 2~Mpc/h from the filament skeleton (see Section \ref{sec:rockstar} for details). In Fig. \ref{fig:subsmpl_env}, we plot the amplitude of the coherence signal as a function of range, for each of the target sub-samples. The coherence amplitude is largest around clusters. The coherence amplitude of groups and filaments is similar, although the group data points are consistently above the filament data points in most distance bins. We clearly see that every amplitude within each structure is greater than amplitude calculated over whole simulation box, since mean density of whole box is lower than density of each structures.

We tested the significance of the correlation between local number density and the amplitude using a Spearman's rank coefficient. But, due to the huge scatter between individual halos, the result was that no significant correlation was detected. This provides further motivation for our approach of stacking all the objects together, and testing the significance of the signal by comparison with multiple catalogues where the spin vectors are randomized in each catalogue. When we take this approach, we find there is indeed a correlation with local number density. For example, see Fig. \ref{fig:hmnd}. Here, the error bars are calculated based on the standard deviation of the amplitude from 100 halo catalogues where, in each catalogue, the halo spin vectors had their directions randomised.

\subsubsection{Host halo mass vs local number densities}

\begin{figure}
    \centering
    \includegraphics[width=\columnwidth]{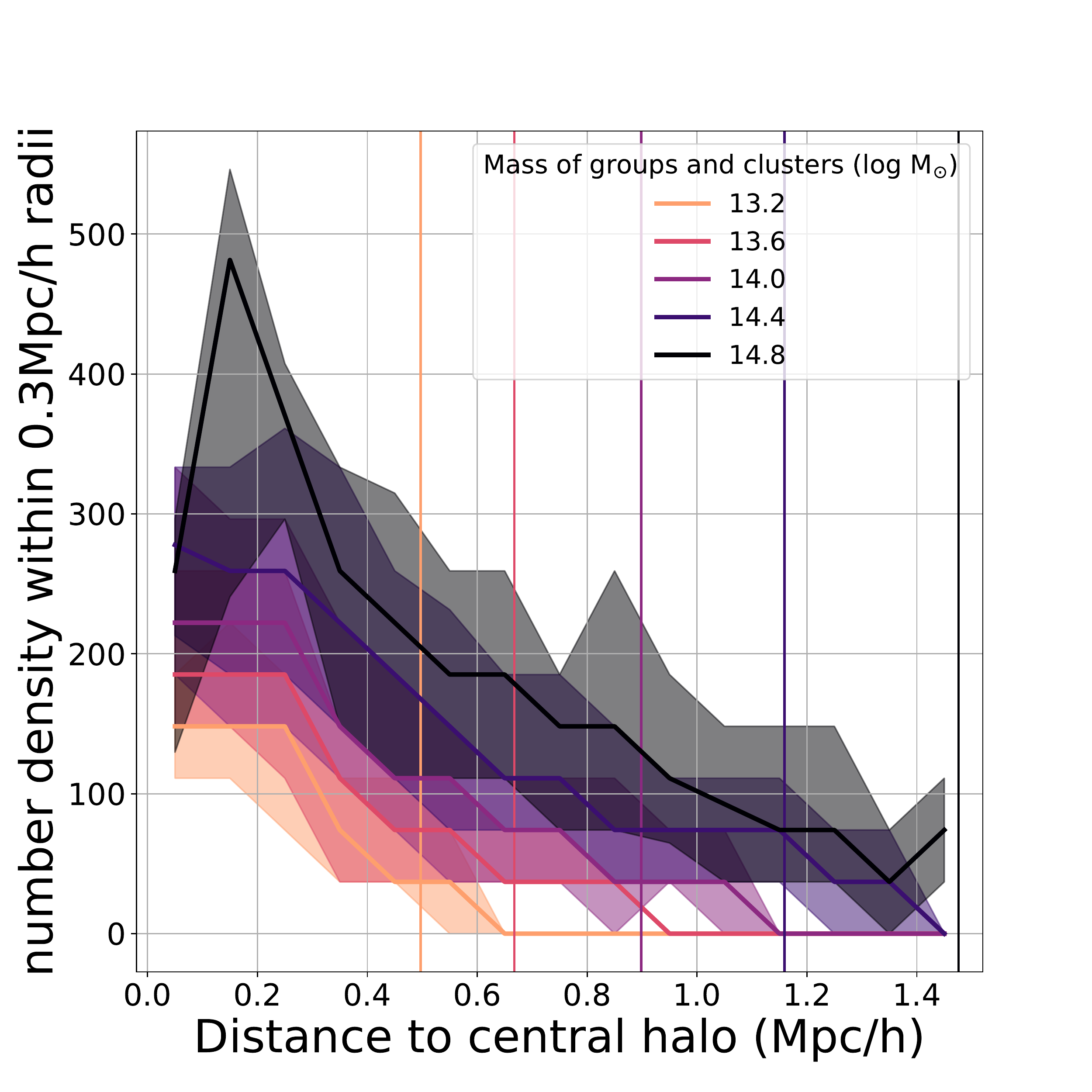}
    \caption{Local number density measured within 300~ kpc/h of halos plotted against distance to the central halo. Solid lines are the mean number densities with shading indicating the quartiles, and there is a different solid line for target halos in different mass hosts from groups up to clusters (see legend). The mean virial radius of central halos in each mass range is plotted as thin colored vertical solid lines.}
    \label{fig:hmndcorr}
\end{figure}

So far, we have seen that the coherence amplitude is larger where the local density is higher, and also in high mass host halos like clusters. However, it is unclear whether host mass or local number density is the primary parameter dictating the coherence amplitude, as more massive clusters might be expected to have higher local number densities near their centers. We confirm this picture in Fig. \ref{fig:hmndcorr} where we plot the local number density as a function of distance from the host center. In this plot, we consider local density within 300~kpc/h instead of the usual 1~Mpc/h. Nevertheless, it confirms that higher mass hosts (see legend for center value of host mass range) indeed have larger local number density at any fixed physical distance. 

\begin{figure}
    \centering
    \includegraphics[width=\columnwidth]{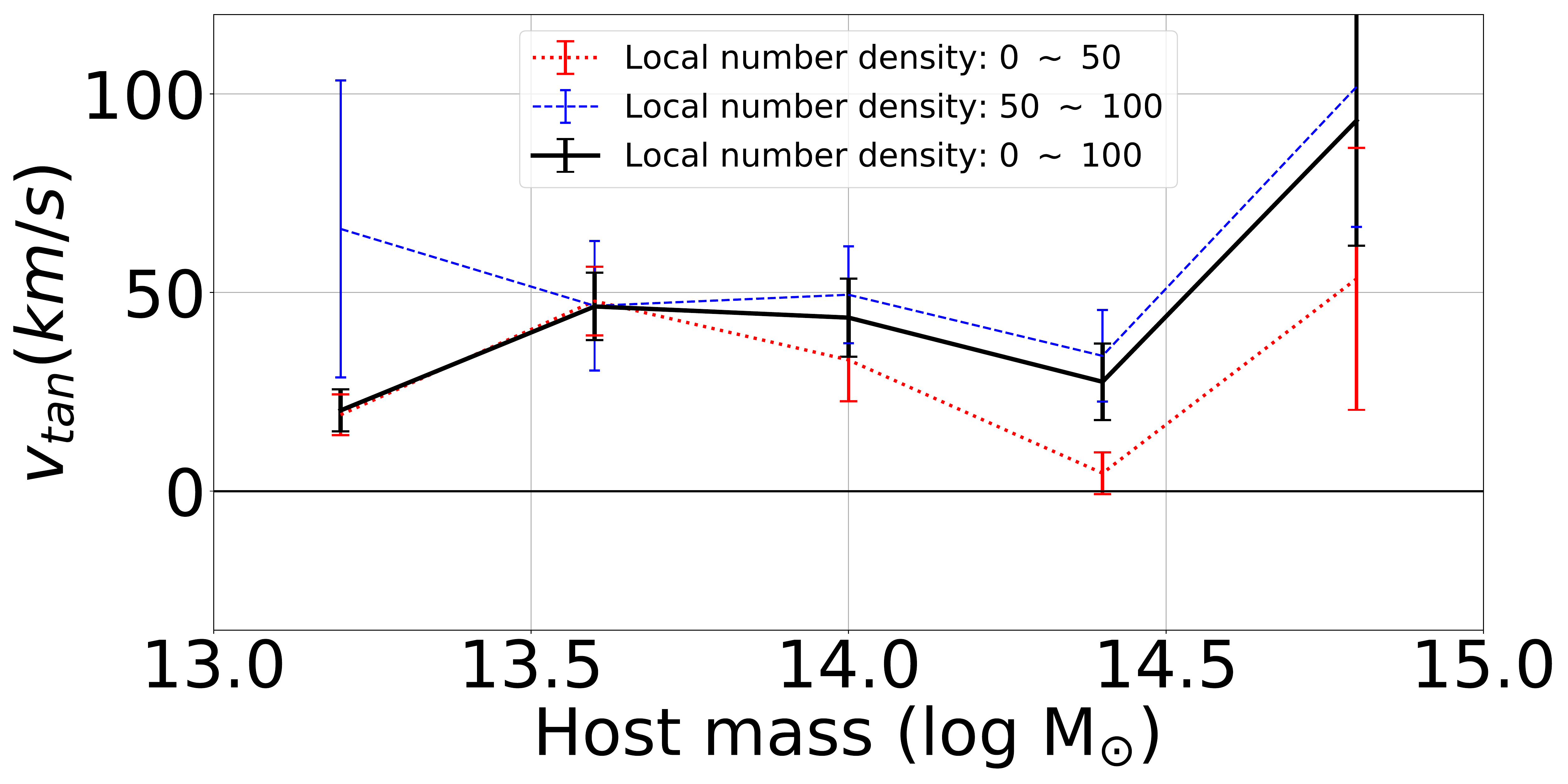}
    \includegraphics[width=\columnwidth]{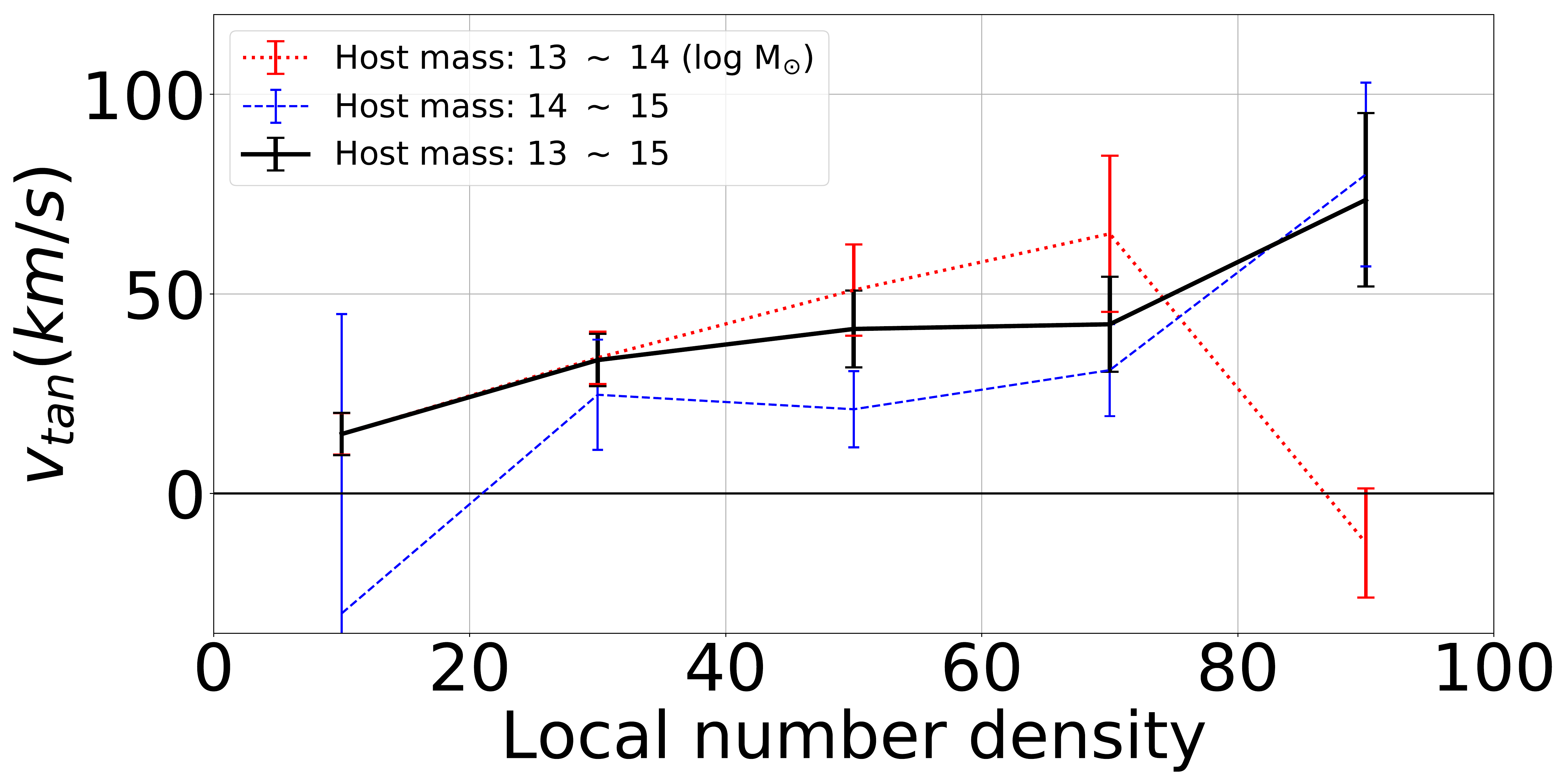}
    \caption{Comparison of the dependency of the coherence amplitude on the mass of host halo (upper panel) and local number density (lower panel) around target halos. Error bars show the standard deviation of the amplitude estimated from 100 randomly spin-oriented catalogs for each 5 boxes. Here, the amplitude of the coherence signal is calculated within 10 virial radius of each host halo. We also divided the halos into subsamples (see legend) as shown with the red and blue lines to attempt to control for the other parameter while varying the parameter on the x-axis. Here, the local density is measured from each halo within a radius of 1 Mpc/h in order to match the integrated result in Fig. \ref{fig:paramdep}}
    \label{fig:hmnd}
\end{figure}

However, the figure also shows there is a range of number density within a particular mass host. Therefore, in the following we try to separate the impact of the two parameters (host mass versus local number density) by varying one parameter while controlling for the other (see Fig. \ref{fig:hmnd}). We do not see a clear correlation between the amplitude of the coherence and host mass (upper panel), for a fixed local number density range (see legend). But, we do see a positive correlation between local number density and the coherence for a fixed host mass range (see lower panel). Thus, the parameter which most strongly influences the coherence amplitude is the local number density rather than the host halo mass. 

\begin{figure}
    \centering
    \includegraphics[width=\columnwidth]{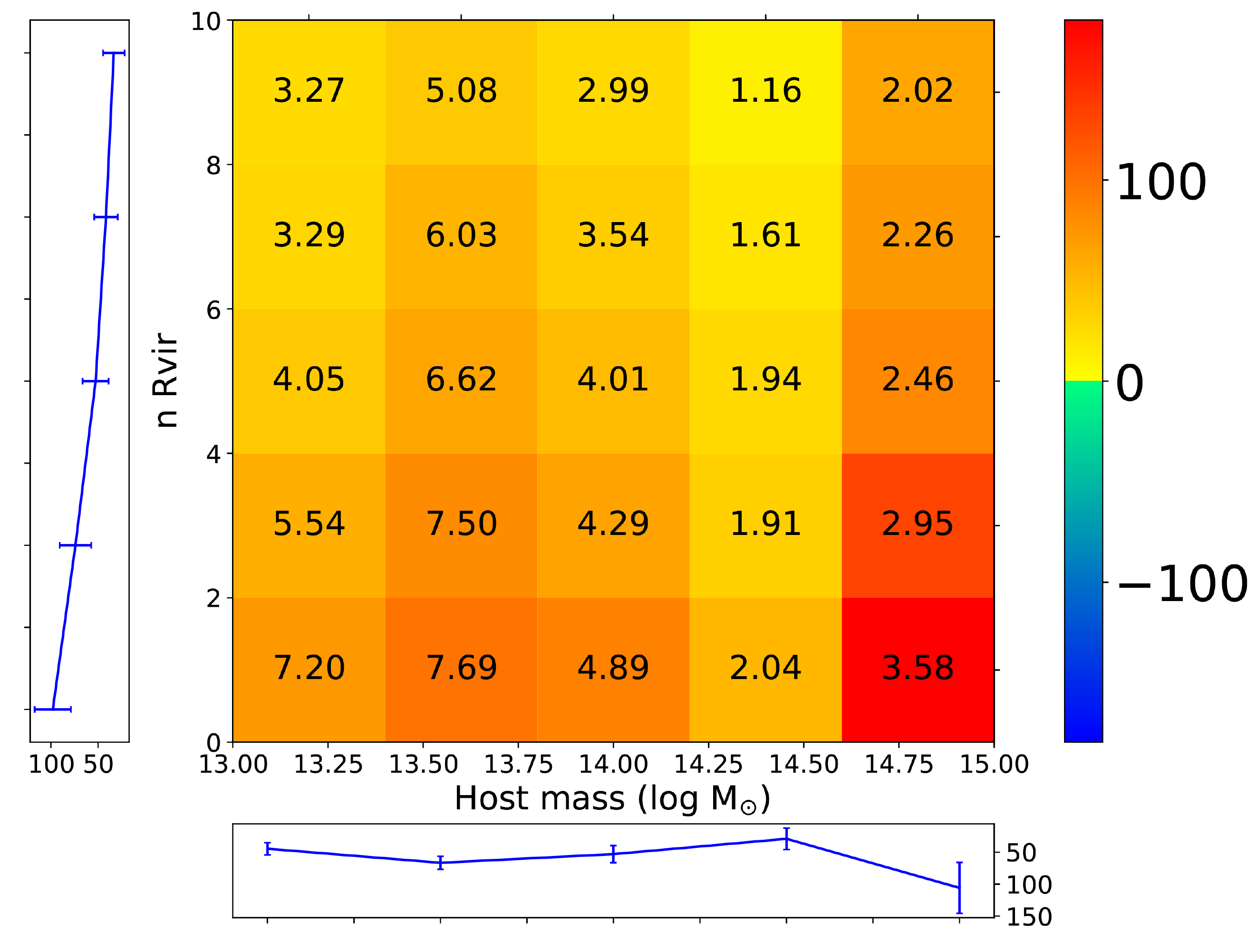}
    \caption{Radial profile of coherence signal amplitude by host mass (x-axis). Figure format is same as Fig. \ref{fig:sampleplot} but cumulative amplitude is plotted along the Y-axis. The total signal amplitude measured for all halos within the given radius is given by the colour bar. The host is the most massive halo containing a target halo. The range of the coherence (y-axis) has been normalized by the virial radius of the host halo.}
    \label{fig:hmdep}
\end{figure}

In other words, the denser inner regions of clusters produce larger amplitudes of coherence signal than the outer regions of the same clusters. In Fig. \ref{fig:hmndcorr} we clearly see the highest number densities are located at smaller distances from central halos. Taking this concept to extremes, in Fig. \ref{fig:hmdep}, we attempt to ascertain the range of the coherence signal for different host masses from groups up to clusters, but this time using a distance scale that is normalised by the virial radius of the host halo (see the y-axis of Fig. \ref{fig:hmdep}). As we saw in the upper panel of Fig. \ref{fig:hmnd}, the dependency on host mass is nearly flat except for the most massive host groups ($>10^{14.5} M_{\odot}$). From the error bars, it can be seen that the significance is highest in the low host mass groups. This might be due to the number statistics of target halos, which are lower in high mass hosts. We count 20039 targets around the lowest mass groups, and only 3278 targets in the most massive clusters. But the distances over which the coherence signal remains detectable can be seen to reach over 10 virial radii from dense systems, which corresponds to nearly 15Mpc/h for the most massive clusters, and the normalised distance is fairly independent of host mass. We note that the local density in Fig. \ref{fig:hmnd} is measured within 1 Mpc/h from each target halo. This is different than the local density measured in Fig. \ref{fig:hmndcorr}, which is measured within 300 kpc/h from the cluster center and is smaller to allow us to measure deeper down into the cluster core.

\subsection{Negative coherence}\label{sec:negacor}
Under specific circumstances, we detect the presence of statistically significant negative coherence in our simulations. Negative coherence means that larger numbers of neighbor halos are orbiting retrograde than prograde to the spin direction of the target halo. In this case, the mean tangential velocity is seen to be negative even when the signal to noise is greater than one. We first saw evidence for negative coherence when we divided our sample by halo mass (see the top row of Fig. \ref{fig:paramdep}.) For long range signal (`3-6~Mpc/h'),  negative coherence is detected in just two cells of the highest mass target halos with a signal to noise of $\sim1.5$. Negative coherence is seen more clearly in the second row of Fig. \ref{fig:paramdep}, when we divide the halos by their spin parameter. In particular, high spin halos tend to show negative coherence with intermediate to long range neighbors. Only when we divide the sample by local number density do we see clear evidence for negative coherence with nearby (0-1~Mpc/h) neighbors (see the final row of Fig. \ref{fig:paramdep}). These cells consistently show high amplitude negative coherence, when the target halo and neighbor halo are embedded in opposing extremes of environmental densities (e.g., target halos in dense cluster cores vs neighbors in low density void surroundings, or vice versa). Although the signal-to-noise is not very high (S/N=1.25--2.54), we do not believe that these negative cells are simply noise. In fact, we see the presence of these negative cells in all 5 simulation volumes separately, and in each case they have similar amplitudes and are found at the same locations in the figures.

\section{Discussion}\label{sec:discussion}

\begin{figure}
    \centering
    \includegraphics[width=\columnwidth]{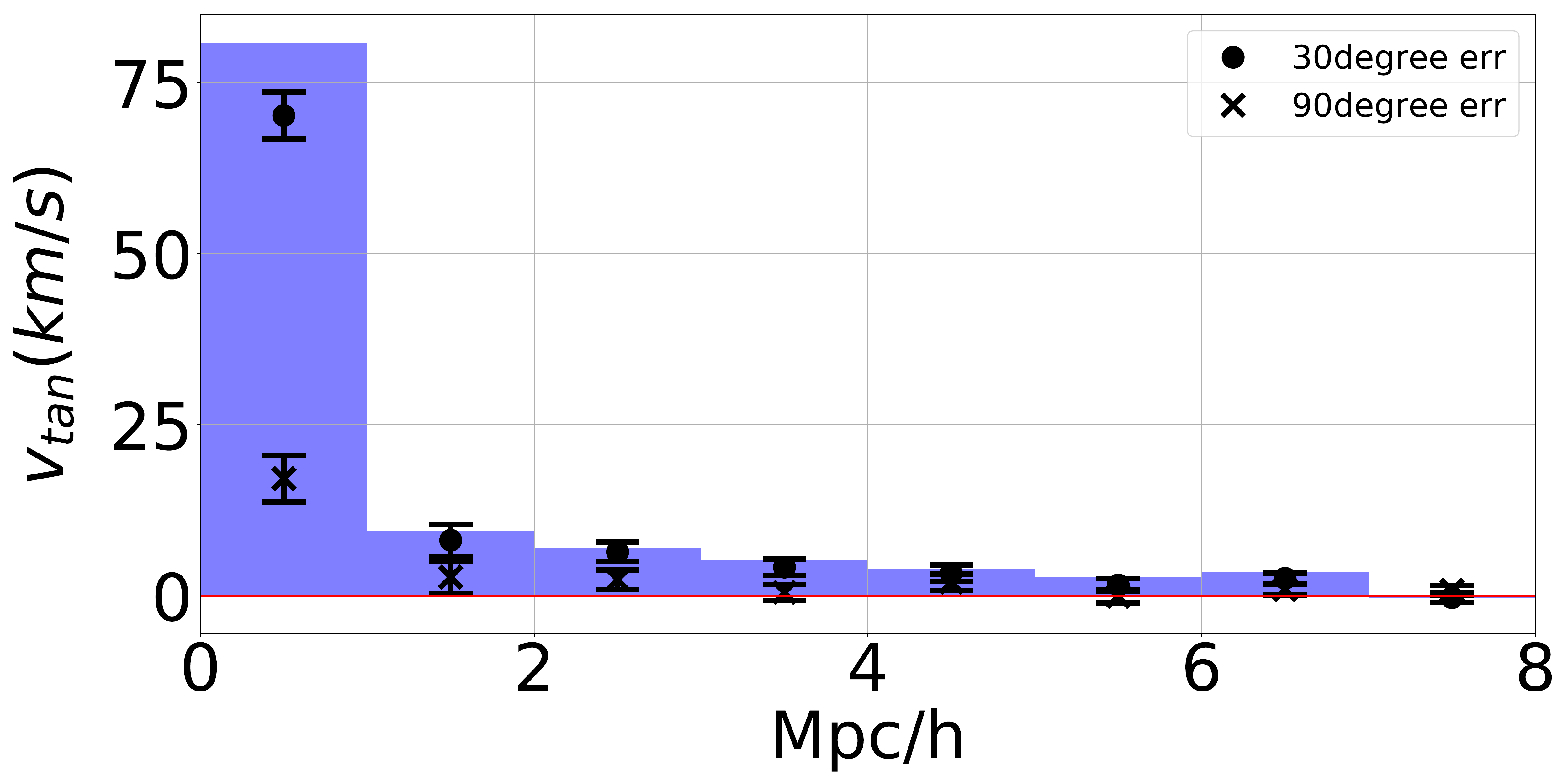}
    \caption{Coherence signal from the first box of DM halo simulation and 30 degree, 90 degree artificial errors. Blue bars represents the original coherence signal, and markers represents corresponding errors. We flipped the target halos spin direction to 100 random directions to achieve the one-sigma errors. The choice of 30 degrees is based on the typical misalignment between DM halos and their stellar disks in cosmological simulations (see the main text for details).}
    \label{fig:errsignal}
\end{figure}

\subsection{On the misalignment between DM halos and galactic discs}\label{sec:misalign}

Given that out simulation study was largely motivated by an observational study, we note that the comparison between the observations of baryonic disks and properties of dark matter halos that we measure is not trivial. We first must consider the fact that the spin vector of DM halos are typically misaligned with the spin of their stellar disks, so they are not perfectly aligned. We chose a 30 degree error based on the typical misalignment that is measured between the DM halo spin and the spin of the stellar disk for massive halos in hydrodynamical cosmological simulations \citep{2015MNRAS.453..721V,2016MNRAS.462.2668T}. We do not know which component would produce the largest amplitude coherence signal. But we can test the impact of giving artificial 30 degrees direction changes to the spin vector of each target halo, and quantify the impact on the coherence amplitude. We assign the 30 degree change in spin vector randomly, and once again generate 100 different random realisations in order to quantify the statistical significance of any changes that are seen. As is shown in Fig. \ref{fig:errsignal}, the addition of a 30 degree misalignment does not strongly reduce the measured coherence amplitudes. In fact, the differences in amplitudes are within one sigma for all distance bins (with the single exception of the 1~Mpc/h bin where it is two sigma different). Meanwhile, the coherence signals were reduced significantly if we injected 90 degrees changes. Interestingly, there is still some weak signal even when we flipped the halo spin by 90 degrees which suggests that some perpendicular coherence might also exist.

\subsection{Comparison with L19b}\label{sec:compare}

In L19b, the coherence amplitude is slightly (0.6-sigma) larger for bright target galaxies at distances between 1~Mpc to 6~Mpc from the target galaxy. For example, the coherence amplitude is $15.9 \pm 24.3$ km/s for faint galaxies ($M_{r}>-20.5$), while the amplitude is $32.6 \pm 14.7$ km/s for bright galaxies ($M_{r} \le -22.5$). Thus it isn't clear if this dependency is reliable. On the contrary, in L19a, which studied close range coherency (within ~800kpc), they report the opposite dependency on halo mass although it is also not very statistically reliable.
In the simulations, we find that the coherence amplitude indeed increases with halo mass within 1~Mpc/h from target halos, but it peaks at approximately 10$^{12.5}$~M$_\odot$ for more distant neighbours. Based on halo abundance matching, we estimate this to be equivalent to a stellar mass of 10$^{10}$~M$_\odot$ for central halos. However, the luminosity of the galaxy would depend on their stellar populations. 

L19b also report a weak, low significance (one-sigma) dependency on local luminosity density, with CALIFA galaxies embedded in denser region presenting slightly higher amplitudes of coherence. However, in the simulations we found a much strong dependency on 3D local number density. Clearly the way in which the environmental density is defined is very different between the observations and our 3D method. Therefore, we decided to mock the observations using the simulation data to better understand why the 3D number density is a more important parameter for the coherence signal.

\begin{figure}
    \centering
    \includegraphics[width=\columnwidth]{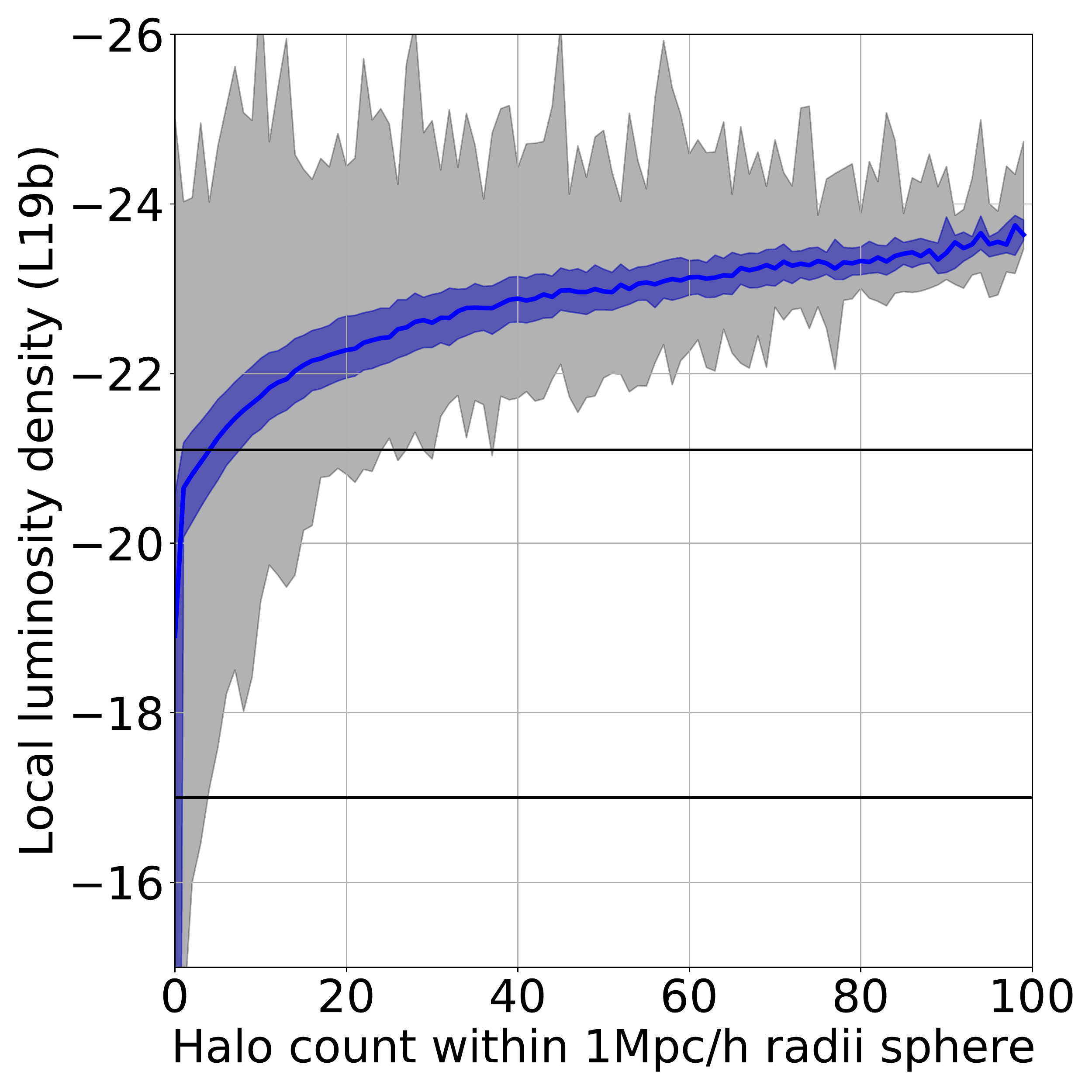}
    \includegraphics[width=\columnwidth]{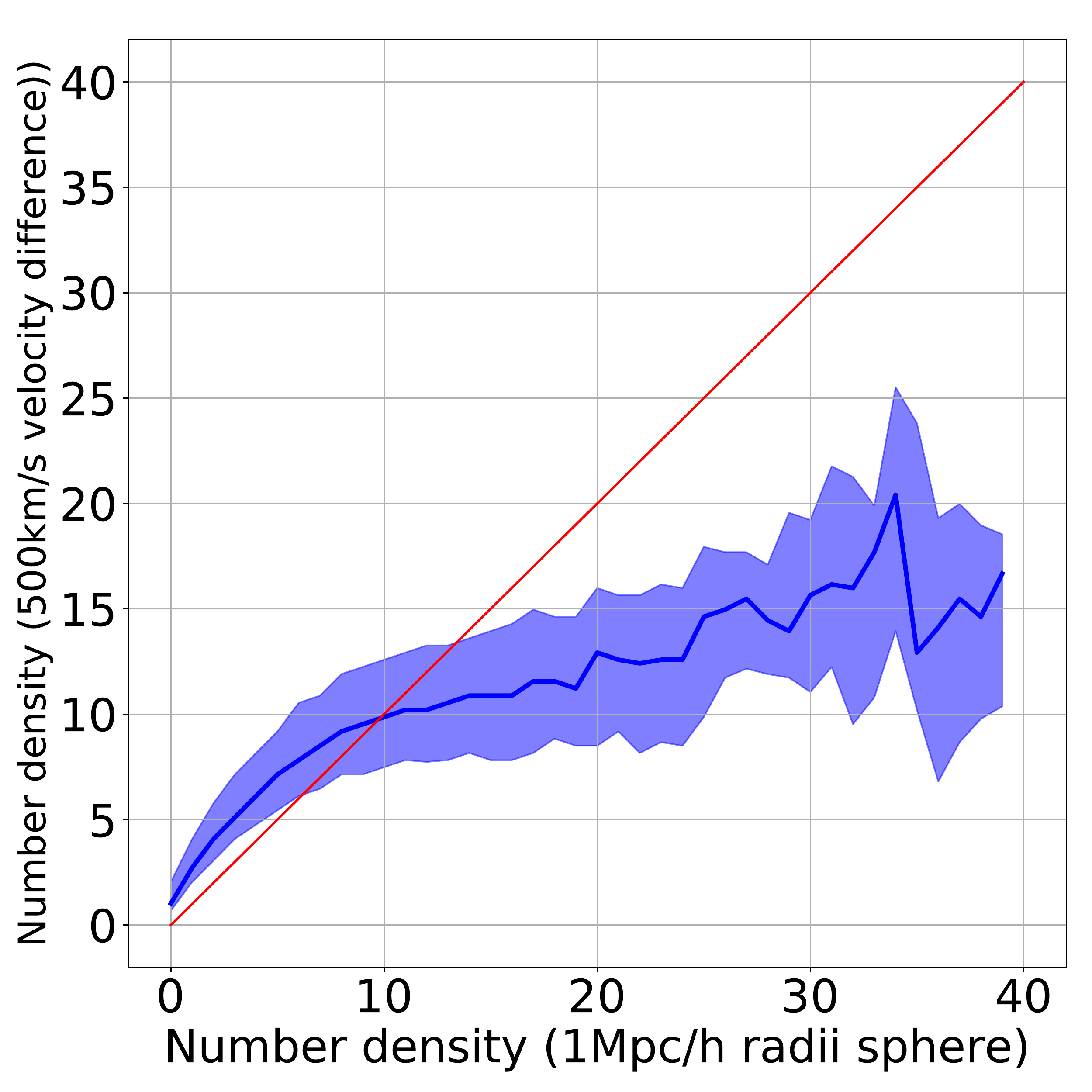}
    \caption{Top: Correlation between number count of halos in our study and local luminosity density suggested by L19b. The grey shaded region is the minimum and maximum data point along the y-axis and the blue shaded region shows the quartiles. Two black solid horizontal lines are correspond to two local luminosity density boundary for dividing galaxies into subsamples in L19b. Bottom: The relation between the number densities in a sphere with 1~Mpc/h radius and a cylinder with a 1~Mpc radius and a 500~km/s line of sight velocity difference. Because of the finger-of-god effect, we do not see a one-to-one relation between these two number densities. The blue line represents the median value and the shaded region shows the quartiles.}
    \label{fig:ndld}
\end{figure}

To estimate the local luminosity density in the simulations, we first estimate the stellar mass for each DM halo using the \citet{2010MNRAS.404.1111G} abundance matching. We then applied a typical stellar mass to light ratio for galaxies in the L19b sample. However, the spread in galaxy color is large at a fixed luminosity, and we do not know the color of a galaxy in our simulations as they are dark matter-only. Therefore, we are forced to naively assume a central value of color from the observed L19b sample of $g-r = 0.7$. Now, using the mass to luminosity relation of \citet{2003ApJS..149..289B} ($log(M/L_{r}) = -(-0.306 + 1.097 * 0.7)$), we applied a constant $M/L_{r}$ of 0.345 for all galaxies.

To mimic the observed dynamics, we consider the halo dynamics down our line-of-sight in the simulation box (in this case, the z-axis). We can now directly adopt the formula for estimating the local luminosity density in L19b. When we compare the model's local luminosity density (y-axis of top panel of Fig. \ref{fig:ndld}) to our 3D local number density (x-axis), it becomes clearer why we do not find the strong dependency that we saw with the 3D local number density. There is a large spread in local luminosity-dependency at fixed 3D local number density, and there is not a strong difference in luminosity density for quite a large range of local number density. 

To attempt to understand why the correlation between local luminosity density and 3D local halo number density is so poor, we tried to remove luminosity effects by instead comparing projected local number density to the 3D local number density. We also restricted the line-of-sight velocity range so that the depth of the volume used to measure the number density would be 1~Mpc if the galaxies are purely in Hubble flow. With the previous L19b velocity range of 500~km/s, the depth of the volume would be much larger ($\sim$5Mpc/h). The results are shown in the bottom panel of Fig. \ref{fig:ndld}. Despite these changes, the high 3D number densities (x-axis) typically do not result in high projected number density (y-axis). Thus it is the deviation from Hubble flow (i.e., the peculiar motions of galaxies in the potential wells of dense environments) that makes it difficult to measure the local number density using a fixed line-of-sight velocity range. These same deviations are behind the well known finger of god (FoG) effect \citep{1972MNRAS.156P...1J, 1987MNRAS.227....1K} which causes galaxies with a high local 3D number density to not have high projected 2D number density in a fixed velocity range. 

These results suggest that a clearer observational dependency on local environment might be revealed if the local density is computed after the FoG effect has been suppressed, under the assumption that the group shape and velocity dispersion is symmetrical both down and perpendicular to our line of sight \citep{2004ApJ...606..702T, 2016ApJ...818..173H, 2020MNRAS.491.4294K}. An additional suggestion for future observations would be to consider both the target and neighbor local number density, as it is the combination of these that more strongly dictates the amplitude of the coherence.

\begin{figure}
    \centering
    \includegraphics[width=\columnwidth]{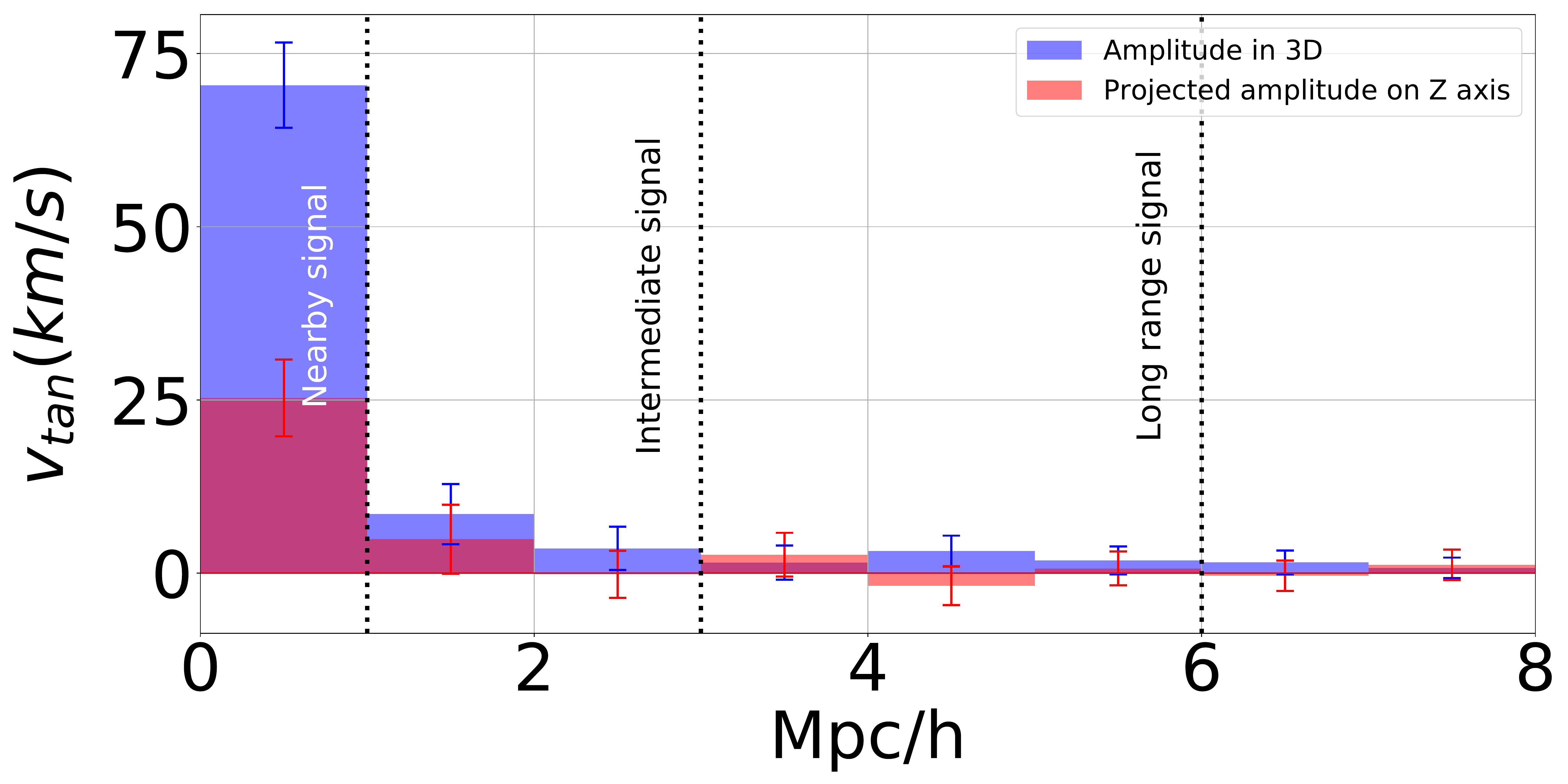}
    \caption{Comparison between the amplitude of coherence measured in 3D and the projected amplitude onto the z-axis of the simulation box. The blue histogram is the same as in Fig. \ref{fig:radial} and the red histogram shows the projected amplitude. Errors are measured in the usual way (randomising the target spin 100 times) but considering projected quiantities only.}
    \label{fig:3d2d}
\end{figure}

So far, we have primarily focused on studying the amplitude of the coherence in three dimensions, rather than the 2D measurements in L19a, L19b. The idea behind this was that we expect to be able to better detect coherence in 3D than in 2D. However, to test how much weaker the signal might be in 2D, we decided to repeat the measurement of the total sample's coherence in 2D, with a line-of-sight along the z-axis, and using quantities projected onto the XY plane. The result is shown in Fig. \ref{fig:3d2d}. As expected, the projected amplitude is about one third of the 3D amplitude in the 1~Mpc/h range and the value is similar with the observed results in L19a (31.1 km/s within 800kpc from target halo), although we struggle to detect the long range coherence in L19b with our projected simulation coherence. For example, L19b find that there is 21~km/s of amplitude at distances up to 6.2~Mpc. In contrast, we find that the amplitude drops in the region where the distance to the target halo is larger than 1~Mpc/h, and it shows less than 3~km/s of projected amplitude within 6~Mpc/h.

We did not find a strong dependency of the coherence amplitude on the spin of halos. L19b found there was more coherence about target galaxies with high spin stellar disks. However, we must keep in mind that differences of coherence between subsamples in L19b are pretty weak, moreover, halo spin and stellar disk spin are not the same quantities. First of all, there could be some misalignment between the two components (see section \ref{sec:misalign} and Fig. \ref{fig:radial2} for more discussion). Secondly, while the magnitude of halo spin and disk spin are expected to be correlated, there is likely some scatter.

We also did not find a strong dependency on the coherence amplitude and the time since the last major merger occurred. It is difficult to compare this result with observations as the time since the last major merger is not easily measured directly in observations. L19b used internal misalignment between the inner ($R<R_e$) and outer ($R_e<R<2*R_e$) stellar disk spin vector as a proxy for merger history dependencies. They found that internally well aligned galaxies have stronger coherence signal despite the lack of dependence that we found, although it is possible that internal misalignment is more sensitive to other properties of the merger history (e.g., rate of mass growth, frequency of minor mergers, gas richness, etc), rather than the single parameter that we considered (i.e., the time since the last major merger).

Our simulations also revealed instances of negative coherence, which was especially strong when the target and neighbor halos were in opposite extremes of local number density (see section \ref{sec:negacor} for a full description). A negative coherence was also reported in L19b, when targets are (1) faint ($M_r > -20.5$), (2) blue ($g-r > 0.756$), (3) diffuse (S$\acute{e}$rsic index $ > 2$) and (4) fast rotators ($(log(L/\left\langle L \right\rangle))-(-0.659M_r-14.144 \ge 0)$ with red neighbors. We do not know the color or how diffuse are the stellar disks of the galaxies in our dark matter-only simulation. But, under the reasonable assumption that fast spin halos contain fast rotation disks, we do see a hint that high spin halos can have show indications of negative coherence (see second row of Fig. \ref{fig:paramdep}).

\subsection{The origins of large scale coherence}\label{sec:scenarios}

Given the known flow of matter from cosmic walls into filaments, and the fact that halo spins tend to align along the filaments \citep{2015MNRAS.452.3369C, 2015MNRAS.446.2744L, 2020arXiv200506479A, 2021NatAs...5..839W, 2021NatAs...5..283M, 2021MNRAS.506.1059X}, it is reasonable to expect that filaments are a likely source of coherence. However, against our expectations, we found that the strongest coherence signals arose in the densest regions of the large scale structure, such as at the centers of clusters and massive groups, and not in filaments (see the third row of Fig. \ref{fig:paramdep} and Fig. \ref{fig:subsmpl_env}).

\begin{figure}
    \centering
    \includegraphics[width=\columnwidth]{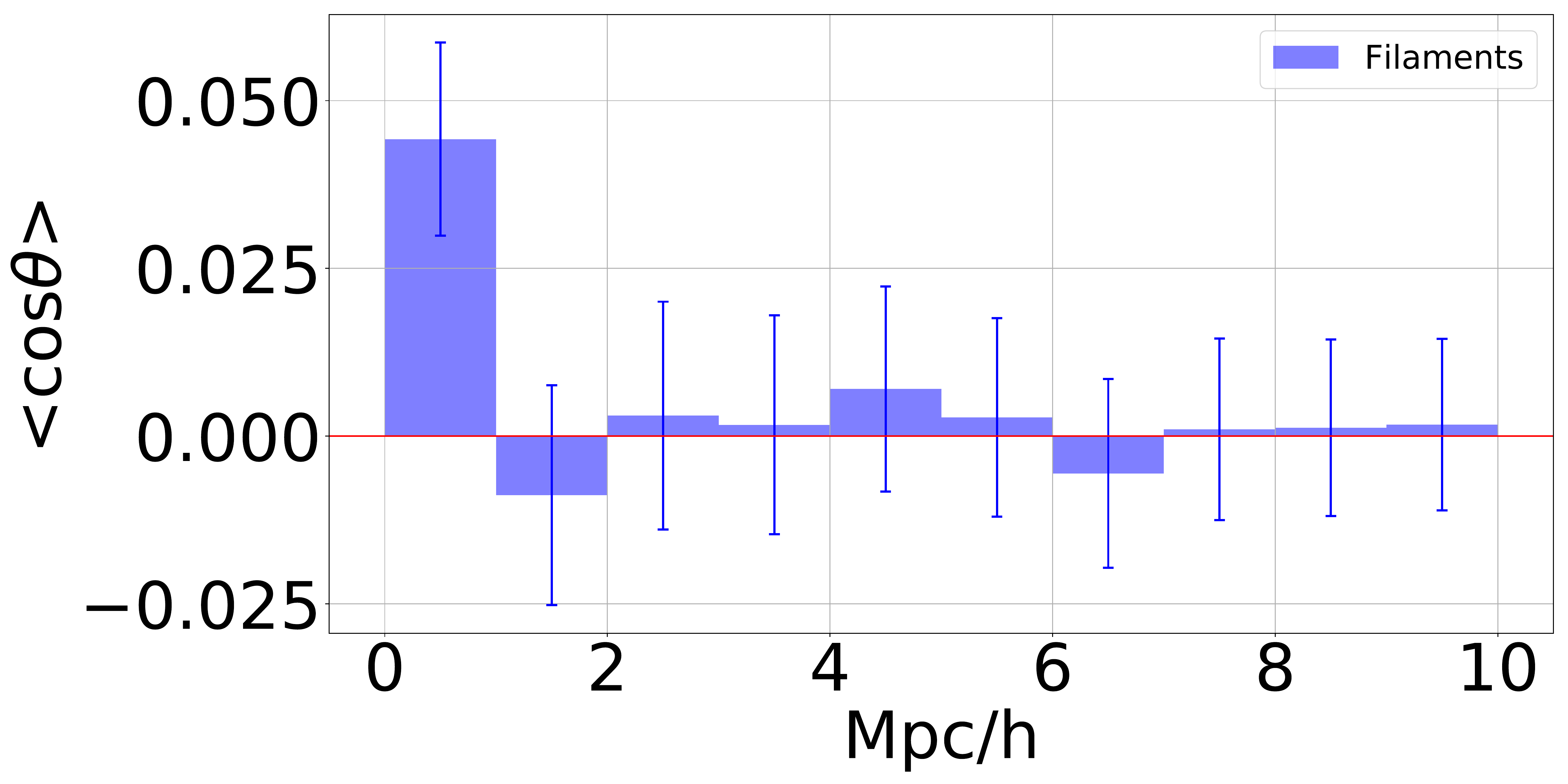}
    \caption{Cosine of angle between halo spin vectors for halos in filaments. Height of each bar represents the mean cosine between the rotation direction of the halo belonging to each group and the rotation direction of whole halos within simulation box at certain range. Errors calculated with 100 halo catalogs in which the target halo's spin direction is randomly orientated.}
    \label{fig:spinspin}
\end{figure}

Furthermore, we also measured the distance over which the spin alignment extends between halos (see Fig. \ref{fig:spinspin}). We find it does not extend over long distances even when we consider halos in filaments, unlike the coherence, perhaps as a result of the bending and twisting of the filament along its length. Nevertheless, this result demonstrates that spin-spin alignment along filament lengths is not a key requirement for large scale coherence. Instead, the coherence can arise with halo spins in more localised regions in the large scale structure.

Although high density environments, such as clusters, seem to be the strongest sources of coherence, the dynamics around filaments could still play a role as clusters form the nodal ends of the filaments. However, most clusters are expected to be connected to more than one filament \citep{2019MNRAS.489.5695D, 2021arXiv210104686G}. And with increasing number of filaments, it is difficult to imagine how the coherence would not be weakened as different filaments intersect the cluster in different directions. Yet, we found that the amplitude of the coherence was significantly stronger in clusters than filaments (see Fig. \ref{fig:subsmpl_env}.)

\begin{figure}
    \centering
    \includegraphics[width=\columnwidth]{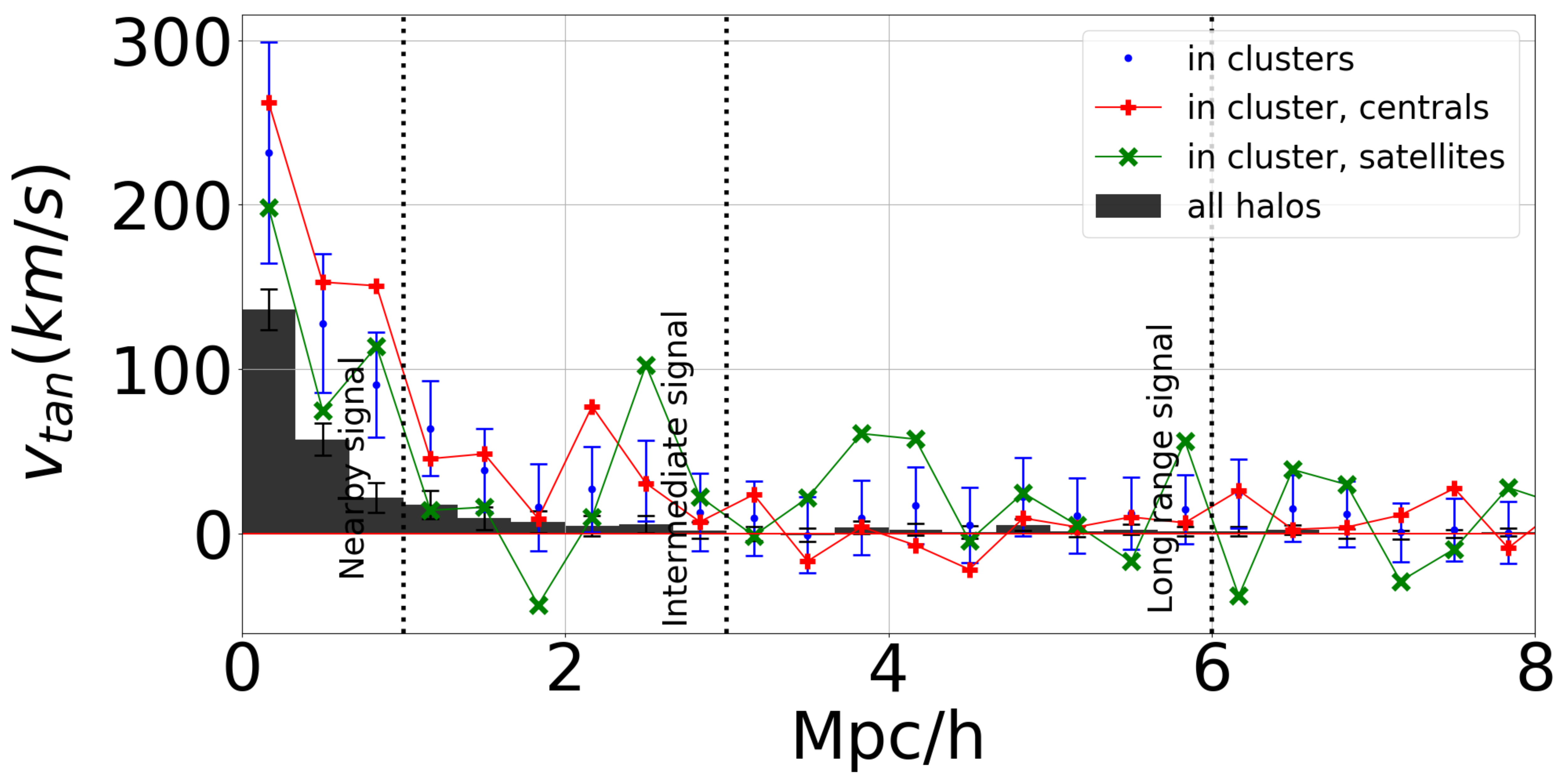}
    \caption{Same as Fig.\ref{fig:subsmpl_env}, but only targets in clusters are used. Also, they are divided into centrals and satellites subsamples so we could compare their amplitudes of coherence.}
    \label{fig:cen_satl}
\end{figure}

An alternative scenario is that the potential well of clusters (or groups) could enhance the coherence in some way. It could perhaps alter the orbits of neighbors, although it is not obvious why they would necessarily become more coherent. Or, the increased rate of mergers in denser environments could act to spin up the halos there, creating enhanced coherence \citep{2015MNRAS.449...49R}. This scenario could make sense for central halos, in which mergers frequently occur. However, once halos become satellites it becomes much more difficult for frequent mergers to occur due to the peculiar motions within the host halo \citep{2010ApJ...715..342H}. Therefore, we might expect a significant difference in the amplitude of the coherence signal between central and satellite halos. We tested this in Fig. \ref{fig:cen_satl}. Central halos in clusters do have a higher amplitude of coherence than satellites in clusters but only at the one-sigma level. This could still be consistent with the enhance spin through mergers scenario if such satellite halos suffered most of their mergers prior to becoming satellites of the cluster, for example in a denser proto-cluster environment. To draw firmer conclusions on this scenario, we believe we would need to study the time evolution of coherence, rather than only considering the $z=0$ snapshot. We also note that we did not see a strong dependency of the coherence amplitude on the scale of the last major merger (see third row of Fig. \ref{fig:paramdep}), which might highlight the importance of minor mergers or smooth accretion in developing the spin of halos rather than specifically requiring major mergers.

The fact that the coherence signal extends out to 6~Mpc/h or more rules out that the long range signal could occur as a result of prior interactions between the target and the neighbors. Any explanation for the origin of coherence must begin with an explanation for how the halo spin was established. Many suggestion suggest that tidal shear from neighboring protohalos is important for establishing the initial halo spin \citep{1951pca..conf..195H, 1969ApJ...155..393P, 1970Ap......6..320D, 1984ApJ...286...38W}. Alternatively, \citet{2013ApJ...766L..15L} suggests that while the early halo spin may have been acquired through tidal torques, at more recent times vorticities in the flows of the large scale structure are primarily responsible for setting the $z=0$ halo spins. The fact that we see large amplitude signals of coherence supports the importance of the vortices in establishing the spin. Otherwise, early halo spins established by tidal torques would have to generate matching vortices that remain fairly constant over the age of the Universe.

Considering our new results and these previous studies, we propose the working theory that vortices in the large scale structure help to set the $z$=0 spin of halos. This could occur by a combination of smooth accretion and mergers, and appears to be especially effective inside the densest regions. We note an almost fractal-like nature to our measured coherence, in that the signal amplitude is nearly constant with host mass (from $10^{13}M_{\odot}$ to $10^{14.5}M_{\odot}$) and with distance (when normalised by host mass virial radius; see Fig. \ref{fig:hmdep}) out to ten virial radii from dense environments. Thus, we picture island-like dense regions, like groups and clusters, each enveloped in their own large scale vortices that extend out to larger physical distances around higher mass clusters.

\section{Summary and Conclusion}\label{sec:conclusion}

We have examined the amplitude of the coherence signal between halo spin and the motions of its neighbors on scales ranging from nearby ($<$1~Mpc/h), to medium range (1-3~Mpc/h), to long range (3-6~Mpc/h) and beyond, using dark matter only cosmological N-body simulations. Our aim was to understand under which circumstances the coherence amplitude is stronger in order to better understand the origins of the phenomena and to give useful feedback to future observational studies of coherence. Our main results are summarized in the following:

(1) Considering all the $z$=0 halos in our simulation combined, a significant signal of coherence was detected. The coherence signal has a high amplitude nearby ($<$1~Mpc/h, Vtan=70~km/s), has decreasing amplitude with increasing distance to neighbors, but remains significant out to 6~Mpc/h.

(2) The mass of target halos plays an important role in controlling the amplitude of the coherence signal. The maximum amplitude of signal is typically for target halos with masses in the $10^{12}M_{\odot}$ to $10^{12.5}M_{\odot}$ range, corresponding to stellar masses of roughly $10^{10}M_{\odot}$. But the target mass where the peak signal amplitude is found decreases as the distance to the neighbors increases.

(3) The local environmental density of halos also plays a key role in dictating the amplitude of the coherence signal for both the target and neighbor halos. High local number densities results in large amplitude coherence signals. Target halos in dense environments, like cluster-mass host halos, have the largest amplitude signal. We confirm that it is primarily the local number density that influences the amplitude of the signal, rather than the host halo mass.

(4) Radial profile of coherence amplitude normalized by virial radius reveal the coherence signal reaches at least 10 virial radius from target halos in dense environments like groups and clusters, which corresponds to striking 10 to 15~Mpc/h in physical distance.

Our results highlight the importance of dense environments such as in clusters and the centers of groups as sources of coherence. While filaments do contribute to the overall coherence, they are not the dominant source in the simulation box (see Fig. \ref{fig:subsmpl_env}). Our results have highlighted some interesting dependencies on halo properties which could give constraints on the true origins of coherence. But we believe the use of high resolution hydrodynamical cosmological simulations, and a study of the time evolution of the coherence amplitudes would provide greater insight in the future. 

Most of all, our results provide useful feedback for observational studies of large scale coherence. In particular, we suggest that the largest amplitudes of coherence would be found in the highest density environments, when the environment of both the target and neighbor halos can be considered simultaneously, using observational measures of the local density that can better suppress the finger-of-god effect. If such measurements are made, we predict that negative amplitude coherence (where neighbors rotate in the opposite direction to their target halo's spin) will be found when target halos and their close neighbors ($<1$Mpc/h) inhabit opposite extremes of the local number density. 

\software{MUSIC\citep{2011MNRAS.415.2101H}, CAMB\citep{2000ApJ...538..473L}, DisPerSE\citep{2011MNRAS.414..350S}, ROCKSTAR\citep{2013ApJ...762..109B}}
\bibliography{bib}{}

\begin{thebibliography}{}
\expandafter\ifx\csname natexlab\endcsname\relax\def\natexlab#1{#1}\fi
\providecommand{\url}[1]{\href{#1}{#1}}
\providecommand{\dodoi}[1]{doi:~\href{http://doi.org/#1}{\nolinkurl{#1}}}
\providecommand{\doeprint}[1]{\href{http://ascl.net/#1}{\nolinkurl{http://ascl.net/#1}}}
\providecommand{\doarXiv}[1]{\href{https://arxiv.org/abs/#1}{\nolinkurl{https://arxiv.org/abs/#1}}}

\bibitem[{{An} {et~al.}(2020){An}, {Kim}, {Moon}, \&
  {Yoon}}]{2020arXiv200506479A}
{An}, S.-H., {Kim}, J., {Moon}, J.-S., \& {Yoon}, S.-J. 2020, arXiv e-prints,
  arXiv:2005.06479.
\newblock \doarXiv{2005.06479}

\bibitem[{{Aragon-Calvo} \& {Yang}(2014)}]{2014MNRAS.440L..46A}
{Aragon-Calvo}, M.~A., \& {Yang}, L.~F. 2014, \mnras, 440, L46,
  \dodoi{10.1093/mnrasl/slu009}

\bibitem[{{Aubert} {et~al.}(2004){Aubert}, {Pichon}, \&
  {Colombi}}]{2004MNRAS.352..376A}
{Aubert}, D., {Pichon}, C., \& {Colombi}, S. 2004, \mnras, 352, 376,
  \dodoi{10.1111/j.1365-2966.2004.07883.x}

\bibitem[{{Bailin} \& {Steinmetz}(2005)}]{2005ApJ...627..647B}
{Bailin}, J., \& {Steinmetz}, M. 2005, \apj, 627, 647, \dodoi{10.1086/430397}

\bibitem[{{Bardeen} {et~al.}(1983){Bardeen}, {Steinhardt}, \&
  {Turner}}]{1983PhRvD..28..679B}
{Bardeen}, J.~M., {Steinhardt}, P.~J., \& {Turner}, M.~S. 1983, \prd, 28, 679,
  \dodoi{10.1103/PhysRevD.28.679}

\bibitem[{{Behroozi} {et~al.}(2013){Behroozi}, {Wechsler}, \&
  {Wu}}]{2013ApJ...762..109B}
{Behroozi}, P.~S., {Wechsler}, R.~H., \& {Wu}, H.-Y. 2013, \apj, 762, 109,
  \dodoi{10.1088/0004-637X/762/2/109}

\bibitem[{{Belfiore} {et~al.}(2015){Belfiore}, {Maiolino}, {Bundy}, {Thomas},
  {Maraston}, {Wilkinson}, {S{\'a}nchez}, {Bershady}, {Blanc}, {Bothwell},
  {Cales}, {Coccato}, {Drory}, {Emsellem}, {Fu}, {Gelfand}, {Law}, {Masters},
  {Parejko}, {Tremonti}, {Wake}, {Weijmans}, {Yan}, {Xiao}, {Zhang}, {Zheng},
  {Bizyaev}, {Kinemuchi}, {Oravetz}, \& {Simmons}}]{2015MNRAS.449..867B}
{Belfiore}, F., {Maiolino}, R., {Bundy}, K., {et~al.} 2015, \mnras, 449, 867,
  \dodoi{10.1093/mnras/stv296}

\bibitem[{{Bell} {et~al.}(2003){Bell}, {McIntosh}, {Katz}, \&
  {Weinberg}}]{2003ApJS..149..289B}
{Bell}, E.~F., {McIntosh}, D.~H., {Katz}, N., \& {Weinberg}, M.~D. 2003, \apjs,
  149, 289, \dodoi{10.1086/378847}

\bibitem[{{Bryant} {et~al.}(2015){Bryant}, {Owers}, {Robotham}, {Croom},
  {Driver}, {Drinkwater}, {Lorente}, {Cortese}, {Scott}, {Colless}, {Schaefer},
  {Taylor}, {Konstantopoulos}, {Allen}, {Baldry}, {Barnes}, {Bauer},
  {Bland-Hawthorn}, {Bloom}, {Brooks}, {Brough}, {Cecil}, {Couch}, {Croton},
  {Davies}, {Ellis}, {Fogarty}, {Foster}, {Glazebrook}, {Goodwin}, {Green},
  {Gunawardhana}, {Hampton}, {Ho}, {Hopkins}, {Kewley}, {Lawrence},
  {Leon-Saval}, {Leslie}, {McElroy}, {Lewis}, {Liske}, {L{\'o}pez-S{\'a}nchez},
  {Mahajan}, {Medling}, {Metcalfe}, {Meyer}, {Mould}, {Obreschkow}, {O'Toole},
  {Pracy}, {Richards}, {Shanks}, {Sharp}, {Sweet}, {Thomas}, {Tonini}, \&
  {Walcher}}]{2015MNRAS.447.2857B}
{Bryant}, J.~J., {Owers}, M.~S., {Robotham}, A.~S.~G., {et~al.} 2015, \mnras,
  447, 2857, \dodoi{10.1093/mnras/stu2635}

\bibitem[{{Bullock} {et~al.}(2001){Bullock}, {Dekel}, {Kolatt}, {Kravtsov},
  {Klypin}, {Porciani}, \& {Primack}}]{2001ApJ...555..240B}
{Bullock}, J.~S., {Dekel}, A., {Kolatt}, T.~S., {et~al.} 2001, \apj, 555, 240,
  \dodoi{10.1086/321477}

\bibitem[{{Bundy} {et~al.}(2015){Bundy}, {Bershady}, {Law}, {Yan}, {Drory},
  {MacDonald}, {Wake}, {Cherinka}, {S{\'a}nchez-Gallego}, {Weijmans}, {Thomas},
  {Tremonti}, {Masters}, {Coccato}, {Diamond-Stanic}, {Arag{\'o}n-Salamanca},
  {Avila-Reese}, {Badenes}, {Falc{\'o}n-Barroso}, {Belfiore}, {Bizyaev},
  {Blanc}, {Bland-Hawthorn}, {Blanton}, {Brownstein}, {Byler}, {Cappellari},
  {Conroy}, {Dutton}, {Emsellem}, {Etherington}, {Frinchaboy}, {Fu}, {Gunn},
  {Harding}, {Johnston}, {Kauffmann}, {Kinemuchi}, {Klaene}, {Knapen},
  {Leauthaud}, {Li}, {Lin}, {Maiolino}, {Malanushenko}, {Malanushenko}, {Mao},
  {Maraston}, {McDermid}, {Merrifield}, {Nichol}, {Oravetz}, {Pan}, {Parejko},
  {Sanchez}, {Schlegel}, {Simmons}, {Steele}, {Steinmetz}, {Thanjavur},
  {Thompson}, {Tinker}, {van den Bosch}, {Westfall}, {Wilkinson}, {Wright},
  {Xiao}, \& {Zhang}}]{2015ApJ...798....7B}
{Bundy}, K., {Bershady}, M.~A., {Law}, D.~R., {et~al.} 2015, \apj, 798, 7,
  \dodoi{10.1088/0004-637X/798/1/7}

\bibitem[{{Cappellari} {et~al.}(2006){Cappellari}, {Bacon}, {Bureau}, {Damen},
  {Davies}, {de Zeeuw}, {Emsellem}, {Falc{\'o}n-Barroso}, {Krajnovi{\'c}},
  {Kuntschner}, {McDermid}, {Peletier}, {Sarzi}, {van den Bosch}, \& {van de
  Ven}}]{2006MNRAS.366.1126C}
{Cappellari}, M., {Bacon}, R., {Bureau}, M., {et~al.} 2006, \mnras, 366, 1126,
  \dodoi{10.1111/j.1365-2966.2005.09981.x}

\bibitem[{{Cappellari} {et~al.}(2011){Cappellari}, {Emsellem}, {Krajnovi{\'c}},
  {McDermid}, {Scott}, {Verdoes Kleijn}, {Young}, {Alatalo}, {Bacon}, {Blitz},
  {Bois}, {Bournaud}, {Bureau}, {Davies}, {Davis}, {de Zeeuw}, {Duc},
  {Khochfar}, {Kuntschner}, {Lablanche}, {Morganti}, {Naab}, {Oosterloo},
  {Sarzi}, {Serra}, \& {Weijmans}}]{2011MNRAS.413..813C}
{Cappellari}, M., {Emsellem}, E., {Krajnovi{\'c}}, D., {et~al.} 2011, \mnras,
  413, 813, \dodoi{10.1111/j.1365-2966.2010.18174.x}

\bibitem[{{Carlberg} {et~al.}(1997){Carlberg}, {Yee}, \&
  {Ellingson}}]{1997ApJ...478..462C}
{Carlberg}, R.~G., {Yee}, H.~K.~C., \& {Ellingson}, E. 1997, \apj, 478, 462,
  \dodoi{10.1086/303805}

\bibitem[{{Codis} {et~al.}(2018){Codis}, {Jindal}, {Chisari}, {Vibert},
  {Dubois}, {Pichon}, \& {Devriendt}}]{2018MNRAS.481.4753C}
{Codis}, S., {Jindal}, A., {Chisari}, N.~E., {et~al.} 2018, \mnras, 481, 4753,
  \dodoi{10.1093/mnras/sty2567}

\bibitem[{{Codis} {et~al.}(2015){Codis}, {Pichon}, \&
  {Pogosyan}}]{2015MNRAS.452.3369C}
{Codis}, S., {Pichon}, C., \& {Pogosyan}, D. 2015, \mnras, 452, 3369,
  \dodoi{10.1093/mnras/stv1570}

\bibitem[{{Colless} {et~al.}(2001){Colless}, {Dalton}, {Maddox}, {Sutherland},
  {Norberg}, {Cole}, {Bland-Hawthorn}, {Bridges}, {Cannon}, {Collins}, {Couch},
  {Cross}, {Deeley}, {De Propris}, {Driver}, {Efstathiou}, {Ellis}, {Frenk},
  {Glazebrook}, {Jackson}, {Lahav}, {Lewis}, {Lumsden}, {Madgwick}, {Peacock},
  {Peterson}, {Price}, {Seaborne}, \& {Taylor}}]{2001MNRAS.328.1039C}
{Colless}, M., {Dalton}, G., {Maddox}, S., {et~al.} 2001, \mnras, 328, 1039,
  \dodoi{10.1046/j.1365-8711.2001.04902.x}

\bibitem[{{Darragh Ford} {et~al.}(2019){Darragh Ford}, {Laigle}, {Gozaliasl},
  {Pichon}, {Devriendt}, {Slyz}, {Arnouts}, {Dubois}, {Finoguenov},
  {Griffiths}, {Kraljic}, {Pan}, {Peirani}, \& {Sarron}}]{2019MNRAS.489.5695D}
{Darragh Ford}, E., {Laigle}, C., {Gozaliasl}, G., {et~al.} 2019, \mnras, 489,
  5695, \dodoi{10.1093/mnras/stz2490}

\bibitem[{{Davis} {et~al.}(1985){Davis}, {Efstathiou}, {Frenk}, \&
  {White}}]{1985ApJ...292..371D}
{Davis}, M., {Efstathiou}, G., {Frenk}, C.~S., \& {White}, S.~D.~M. 1985, \apj,
  292, 371, \dodoi{10.1086/163168}

\bibitem[{{Davis} {et~al.}(1982){Davis}, {Huchra}, {Latham}, \&
  {Tonry}}]{1982ApJ...253..423D}
{Davis}, M., {Huchra}, J., {Latham}, D.~W., \& {Tonry}, J. 1982, \apj, 253,
  423, \dodoi{10.1086/159646}

\bibitem[{{Doroshkevich}(1970)}]{1970Ap......6..320D}
{Doroshkevich}, A.~G. 1970, Astrophysics, 6, 320, \dodoi{10.1007/BF01001625}

\bibitem[{{Emsellem} {et~al.}(2007){Emsellem}, {Cappellari}, {Krajnovi{\'c}},
  {van de Ven}, {Bacon}, {Bureau}, {Davies}, {de Zeeuw}, {Falc{\'o}n-Barroso},
  {Kuntschner}, {McDermid}, {Peletier}, \& {Sarzi}}]{2007MNRAS.379..401E}
{Emsellem}, E., {Cappellari}, M., {Krajnovi{\'c}}, D., {et~al.} 2007, \mnras,
  379, 401, \dodoi{10.1111/j.1365-2966.2007.11752.x}

\bibitem[{{Emsellem} {et~al.}(2011){Emsellem}, {Cappellari}, {Krajnovi{\'c}},
  {Alatalo}, {Blitz}, {Bois}, {Bournaud}, {Bureau}, {Davies}, {Davis}, {de
  Zeeuw}, {Khochfar}, {Kuntschner}, {Lablanche}, {McDermid}, {Morganti},
  {Naab}, {Oosterloo}, {Sarzi}, {Scott}, {Serra}, {van de Ven}, {Weijmans}, \&
  {Young}}]{2011MNRAS.414..888E}
---. 2011, \mnras, 414, 888, \dodoi{10.1111/j.1365-2966.2011.18496.x}

\bibitem[{{Falco} {et~al.}(1999){Falco}, {Kurtz}, {Geller}, {Huchra}, {Peters},
  {Berlind}, {Mink}, {Tokarz}, \& {Elwell}}]{1999PASP..111..438F}
{Falco}, E.~E., {Kurtz}, M.~J., {Geller}, M.~J., {et~al.} 1999, \pasp, 111,
  438, \dodoi{10.1086/316343}

\bibitem[{{Girardi} {et~al.}(1995){Girardi}, {Biviano}, {Giuricin},
  {Mardirossian}, \& {Mezzetti}}]{1995ApJ...438..527G}
{Girardi}, M., {Biviano}, A., {Giuricin}, G., {Mardirossian}, F., \&
  {Mezzetti}, M. 1995, \apj, 438, 527, \dodoi{10.1086/175099}

\bibitem[{{Goddard} {et~al.}(2017){Goddard}, {Thomas}, {Maraston}, {Westfall},
  {Etherington}, {Riffel}, {Mallmann}, {Zheng}, {Argudo-Fern{\'a}ndez},
  {Bershady}, {Bundy}, {Drory}, {Law}, {Yan}, {Wake}, {Weijmans}, {Bizyaev},
  {Brownstein}, {Lane}, {Maiolino}, {Masters}, {Merrifield}, {Nitschelm},
  {Pan}, {Roman-Lopes}, \& {Storchi-Bergmann}}]{2017MNRAS.465..688G}
{Goddard}, D., {Thomas}, D., {Maraston}, C., {et~al.} 2017, \mnras, 465, 688,
  \dodoi{10.1093/mnras/stw2719}

\bibitem[{{Gott} {et~al.}(2005){Gott}, {Juri{\'c}}, {Schlegel}, {Hoyle},
  {Vogeley}, {Tegmark}, {Bahcall}, \& {Brinkmann}}]{2005ApJ...624..463G}
{Gott}, J.~Richard, I., {Juri{\'c}}, M., {Schlegel}, D., {et~al.} 2005, \apj,
  624, 463, \dodoi{10.1086/428890}

\bibitem[{{Gott} {et~al.}(1986){Gott}, {Melott}, \&
  {Dickinson}}]{1986ApJ...306..341G}
{Gott}, J.~Richard, I., {Melott}, A.~L., \& {Dickinson}, M. 1986, \apj, 306,
  341, \dodoi{10.1086/164347}

\bibitem[{{Gouin} {et~al.}(2021){Gouin}, {Bonnaire}, \&
  {Aghanim}}]{2021arXiv210104686G}
{Gouin}, C., {Bonnaire}, T., \& {Aghanim}, N. 2021, arXiv e-prints,
  arXiv:2101.04686.
\newblock \doarXiv{2101.04686}

\bibitem[{{Guo} {et~al.}(2010){Guo}, {White}, {Li}, \&
  {Boylan-Kolchin}}]{2010MNRAS.404.1111G}
{Guo}, Q., {White}, S., {Li}, C., \& {Boylan-Kolchin}, M. 2010, \mnras, 404,
  1111, \dodoi{10.1111/j.1365-2966.2010.16341.x}

\bibitem[{{Hahn} \& {Abel}(2011)}]{2011MNRAS.415.2101H}
{Hahn}, O., \& {Abel}, T. 2011, \mnras, 415, 2101,
  \dodoi{10.1111/j.1365-2966.2011.18820.x}

\bibitem[{{Hahn} {et~al.}(2007){Hahn}, {Carollo}, {Porciani}, \&
  {Dekel}}]{2007MNRAS.381...41H}
{Hahn}, O., {Carollo}, C.~M., {Porciani}, C., \& {Dekel}, A. 2007, \mnras, 381,
  41, \dodoi{10.1111/j.1365-2966.2007.12249.x}

\bibitem[{{Hawking}(1982)}]{1982PhLB..115..295H}
{Hawking}, S.~W. 1982, Physics Letters B, 115, 295,
  \dodoi{10.1016/0370-2693(82)90373-2}

\bibitem[{{Hester} \& {Tasitsiomi}(2010)}]{2010ApJ...715..342H}
{Hester}, J.~A., \& {Tasitsiomi}, A. 2010, \apj, 715, 342,
  \dodoi{10.1088/0004-637X/715/1/342}

\bibitem[{{Hoyle}(1951)}]{1951pca..conf..195H}
{Hoyle}, F. 1951, in Problems of Cosmical Aerodynamics, 195

\bibitem[{{Huchra} {et~al.}(2012){Huchra}, {Macri}, {Masters}, {Jarrett},
  {Berlind}, {Calkins}, {Crook}, {Cutri}, {Erdo{\v{g}}du}, {Falco}, {George},
  {Hutcheson}, {Lahav}, {Mader}, {Mink}, {Martimbeau}, {Schneider},
  {Skrutskie}, {Tokarz}, \& {Westover}}]{2012ApJS..199...26H}
{Huchra}, J.~P., {Macri}, L.~M., {Masters}, K.~L., {et~al.} 2012, \apjs, 199,
  26, \dodoi{10.1088/0067-0049/199/2/26}

\bibitem[{{Hwang} {et~al.}(2016){Hwang}, {Geller}, {Park}, {Fabricant},
  {Kurtz}, {Rines}, {Kim}, {Diaferio}, {Zahid}, {Berlind}, {Calkins}, {Tokarz},
  \& {Moran}}]{2016ApJ...818..173H}
{Hwang}, H.~S., {Geller}, M.~J., {Park}, C., {et~al.} 2016, \apj, 818, 173,
  \dodoi{10.3847/0004-637X/818/2/173}

\bibitem[{{Ibarra-Medel} {et~al.}(2016){Ibarra-Medel}, {S{\'a}nchez},
  {Avila-Reese}, {Hern{\'a}ndez-Toledo}, {Gonz{\'a}lez}, {Drory}, {Bundy},
  {Bizyaev}, {Cano-D{\'\i}az}, {Malanushenko}, {Pan}, {Roman-Lopes}, \&
  {Thomas}}]{2016MNRAS.463.2799I}
{Ibarra-Medel}, H.~J., {S{\'a}nchez}, S.~F., {Avila-Reese}, V., {et~al.} 2016,
  \mnras, 463, 2799, \dodoi{10.1093/mnras/stw2126}

\bibitem[{{Jackson}(1972)}]{1972MNRAS.156P...1J}
{Jackson}, J.~C. 1972, \mnras, 156, 1P, \dodoi{10.1093/mnras/156.1.1P}

\bibitem[{{Jhee} {et~al.}(2022){Jhee}, {Song}, {Smith}, {Shin}, {Park}, \&
  {Laigle}}]{2022arXiv220109540J}
{Jhee}, H., {Song}, H., {Smith}, R., {et~al.} 2022, arXiv e-prints,
  arXiv:2201.09540.
\newblock \doarXiv{2201.09540}

\bibitem[{{Jones} {et~al.}(2009){Jones}, {Read}, {Saunders}, {Colless},
  {Jarrett}, {Parker}, {Fairall}, {Mauch}, {Sadler}, {Watson}, {Burton},
  {Campbell}, {Cass}, {Croom}, {Dawe}, {Fiegert}, {Frankcombe}, {Hartley},
  {Huchra}, {James}, {Kirby}, {Lahav}, {Lucey}, {Mamon}, {Moore}, {Peterson},
  {Prior}, {Proust}, {Russell}, {Safouris}, {Wakamatsu}, {Westra}, \&
  {Williams}}]{2009MNRAS.399..683J}
{Jones}, D.~H., {Read}, M.~A., {Saunders}, W., {et~al.} 2009, \mnras, 399, 683,
  \dodoi{10.1111/j.1365-2966.2009.15338.x}

\bibitem[{{Kaiser}(1987)}]{1987MNRAS.227....1K}
{Kaiser}, N. 1987, \mnras, 227, 1, \dodoi{10.1093/mnras/227.1.1}

\bibitem[{{Krajnovi{\'c}} {et~al.}(2011){Krajnovi{\'c}}, {Emsellem},
  {Cappellari}, {Alatalo}, {Blitz}, {Bois}, {Bournaud}, {Bureau}, {Davies},
  {Davis}, {de Zeeuw}, {Khochfar}, {Kuntschner}, {Lablanche}, {McDermid},
  {Morganti}, {Naab}, {Oosterloo}, {Sarzi}, {Scott}, {Serra}, {Weijmans}, \&
  {Young}}]{2011MNRAS.414.2923K}
{Krajnovi{\'c}}, D., {Emsellem}, E., {Cappellari}, M., {et~al.} 2011, \mnras,
  414, 2923, \dodoi{10.1111/j.1365-2966.2011.18560.x}

\bibitem[{{Kraljic} {et~al.}(2020){Kraljic}, {Pichon}, {Codis}, {Laigle},
  {Dav{\'e}}, {Dubois}, {Hwang}, {Pogosyan}, {Arnouts}, {Devriendt}, {Musso},
  {Peirani}, {Slyz}, \& {Treyer}}]{2020MNRAS.491.4294K}
{Kraljic}, K., {Pichon}, C., {Codis}, S., {et~al.} 2020, \mnras, 491, 4294,
  \dodoi{10.1093/mnras/stz3319}

\bibitem[{{Laigle} {et~al.}(2015){Laigle}, {Pichon}, {Codis}, {Dubois}, {Le
  Borgne}, {Pogosyan}, {Devriendt}, {Peirani}, {Prunet}, {Rouberol}, {Slyz}, \&
  {Sousbie}}]{2015MNRAS.446.2744L}
{Laigle}, C., {Pichon}, C., {Codis}, S., {et~al.} 2015, \mnras, 446, 2744,
  \dodoi{10.1093/mnras/stu2289}

\bibitem[{{Lee} {et~al.}(2019{\natexlab{a}}){Lee}, {Pak}, {Lee}, \&
  {Song}}]{2019ApJ...872...78L}
{Lee}, J.~H., {Pak}, M., {Lee}, H.-R., \& {Song}, H. 2019{\natexlab{a}}, \apj,
  872, 78, \dodoi{10.3847/1538-4357/aafcb4}

\bibitem[{{Lee} {et~al.}(2019{\natexlab{b}}){Lee}, {Pak}, {Song}, {Lee}, {Kim},
  \& {Jeong}}]{2019ApJ...884..104L}
{Lee}, J.~H., {Pak}, M., {Song}, H., {et~al.} 2019{\natexlab{b}}, \apj, 884,
  104, \dodoi{10.3847/1538-4357/ab3fa3}

\bibitem[{{Lewis} {et~al.}(2000){Lewis}, {Challinor}, \&
  {Lasenby}}]{2000ApJ...538..473L}
{Lewis}, A., {Challinor}, A., \& {Lasenby}, A. 2000, \apj, 538, 473,
  \dodoi{10.1086/309179}

\bibitem[{{Libeskind} {et~al.}(2012){Libeskind}, {Hoffman}, {Knebe},
  {Steinmetz}, {Gottl{\"o}ber}, {Metuki}, \& {Yepes}}]{2012MNRAS.421L.137L}
{Libeskind}, N.~I., {Hoffman}, Y., {Knebe}, A., {et~al.} 2012, \mnras, 421,
  L137, \dodoi{10.1111/j.1745-3933.2012.01222.x}

\bibitem[{{Libeskind} {et~al.}(2013){Libeskind}, {Hoffman}, {Steinmetz},
  {Gottl{\"o}ber}, {Knebe}, \& {Hess}}]{2013ApJ...766L..15L}
{Libeskind}, N.~I., {Hoffman}, Y., {Steinmetz}, M., {et~al.} 2013, \apjl, 766,
  L15, \dodoi{10.1088/2041-8205/766/2/L15}

\bibitem[{{Motloch} {et~al.}(2021){Motloch}, {Yu}, {Pen}, \&
  {Xie}}]{2021NatAs...5..283M}
{Motloch}, P., {Yu}, H.-R., {Pen}, U.-L., \& {Xie}, Y. 2021, Nature Astronomy,
  5, 283, \dodoi{10.1038/s41550-020-01262-3}

\bibitem[{{Paz} {et~al.}(2008){Paz}, {Stasyszyn}, \&
  {Padilla}}]{2008MNRAS.389.1127P}
{Paz}, D.~J., {Stasyszyn}, F., \& {Padilla}, N.~D. 2008, \mnras, 389, 1127,
  \dodoi{10.1111/j.1365-2966.2008.13655.x}

\bibitem[{{Peebles}(1969)}]{1969ApJ...155..393P}
{Peebles}, P.~J.~E. 1969, \apj, 155, 393, \dodoi{10.1086/149876}

\bibitem[{{Rodriguez-Gomez} {et~al.}(2015){Rodriguez-Gomez}, {Genel},
  {Vogelsberger}, {Sijacki}, {Pillepich}, {Sales}, {Torrey}, {Snyder},
  {Nelson}, {Springel}, {Ma}, \& {Hernquist}}]{2015MNRAS.449...49R}
{Rodriguez-Gomez}, V., {Genel}, S., {Vogelsberger}, M., {et~al.} 2015, \mnras,
  449, 49, \dodoi{10.1093/mnras/stv264}

\bibitem[{{S{\'a}nchez} {et~al.}(2012){S{\'a}nchez}, {Kennicutt}, {Gil de Paz},
  {van de Ven}, {V{\'\i}lchez}, {Wisotzki}, {Walcher}, {Mast}, {Aguerri},
  {Albiol-P{\'e}rez}, {Alonso-Herrero}, {Alves}, {Bakos}, {Bart{\'a}kov{\'a}},
  {Bland-Hawthorn}, {Boselli}, {Bomans}, {Castillo-Morales}, {Cortijo-Ferrero},
  {de Lorenzo-C{\'a}ceres}, {Del Olmo}, {Dettmar}, {D{\'\i}az}, {Ellis},
  {Falc{\'o}n-Barroso}, {Flores}, {Gallazzi}, {Garc{\'\i}a-Lorenzo},
  {Gonz{\'a}lez Delgado}, {Gruel}, {Haines}, {Hao}, {Husemann},
  {Igl{\'e}sias-P{\'a}ramo}, {Jahnke}, {Johnson}, {Jungwiert}, {Kalinova},
  {Kehrig}, {Kupko}, {L{\'o}pez-S{\'a}nchez}, {Lyubenova}, {Marino},
  {M{\'a}rmol-Queralt{\'o}}, {M{\'a}rquez}, {Masegosa}, {Meidt},
  {Mendez-Abreu}, {Monreal-Ibero}, {Montijo}, {Mour{\~a}o}, {Palacios-Navarro},
  {Papaderos}, {Pasquali}, {Peletier}, {P{\'e}rez}, {P{\'e}rez}, {Quirrenbach},
  {Rela{\~n}o}, {Rosales-Ortega}, {Roth}, {Ruiz-Lara},
  {S{\'a}nchez-Bl{\'a}zquez}, {Sengupta}, {Singh}, {Stanishev}, {Trager},
  {Vazdekis}, {Viironen}, {Wild}, {Zibetti}, \&
  {Ziegler}}]{2012A&A...538A...8S}
{S{\'a}nchez}, S.~F., {Kennicutt}, R.~C., {Gil de Paz}, A., {et~al.} 2012,
  \aap, 538, A8, \dodoi{10.1051/0004-6361/201117353}

\bibitem[{{Sousbie}(2011)}]{2011MNRAS.414..350S}
{Sousbie}, T. 2011, \mnras, 414, 350, \dodoi{10.1111/j.1365-2966.2011.18394.x}

\bibitem[{{Springel} {et~al.}(2001){Springel}, {Yoshida}, \&
  {White}}]{2001NewA....6...79S}
{Springel}, V., {Yoshida}, N., \& {White}, S. D.~M. 2001, \na, 6, 79,
  \dodoi{10.1016/S1384-1076(01)00042-2}

\bibitem[{{Starobinsky}(1982)}]{1982PhLB..117..175S}
{Starobinsky}, A.~A. 1982, Physics Letters B, 117, 175,
  \dodoi{10.1016/0370-2693(82)90541-X}

\bibitem[{{Tegmark} {et~al.}(2004){Tegmark}, {Blanton}, {Strauss}, {Hoyle},
  {Schlegel}, {Scoccimarro}, {Vogeley}, {Weinberg}, {Zehavi}, {Berlind},
  {Budavari}, {Connolly}, {Eisenstein}, {Finkbeiner}, {Frieman}, {Gunn},
  {Hamilton}, {Hui}, {Jain}, {Johnston}, {Kent}, {Lin}, {Nakajima}, {Nichol},
  {Ostriker}, {Pope}, {Scranton}, {Seljak}, {Sheth}, {Stebbins}, {Szalay},
  {Szapudi}, {Verde}, {Xu}, {Annis}, {Bahcall}, {Brinkmann}, {Burles},
  {Castander}, {Csabai}, {Loveday}, {Doi}, {Fukugita}, {Gott}, {Hennessy},
  {Hogg}, {Ivezi{\'c}}, {Knapp}, {Lamb}, {Lee}, {Lupton}, {McKay}, {Kunszt},
  {Munn}, {O'Connell}, {Peoples}, {Pier}, {Richmond}, {Rockosi}, {Schneider},
  {Stoughton}, {Tucker}, {Vanden Berk}, {Yanny}, {York}, \& {SDSS
  Collaboration}}]{2004ApJ...606..702T}
{Tegmark}, M., {Blanton}, M.~R., {Strauss}, M.~A., {et~al.} 2004, \apj, 606,
  702, \dodoi{10.1086/382125}

\bibitem[{{Tenneti} {et~al.}(2016){Tenneti}, {Mandelbaum}, \& {Di
  Matteo}}]{2016MNRAS.462.2668T}
{Tenneti}, A., {Mandelbaum}, R., \& {Di Matteo}, T. 2016, \mnras, 462, 2668,
  \dodoi{10.1093/mnras/stw1823}

\bibitem[{{Trowland} {et~al.}(2013){Trowland}, {Lewis}, \&
  {Bland-Hawthorn}}]{2013ApJ...762...72T}
{Trowland}, H.~E., {Lewis}, G.~F., \& {Bland-Hawthorn}, J. 2013, \apj, 762, 72,
  \dodoi{10.1088/0004-637X/762/2/72}

\bibitem[{{Velliscig} {et~al.}(2015){Velliscig}, {Cacciato}, {Schaye}, {Crain},
  {Bower}, {van Daalen}, {Dalla Vecchia}, {Frenk}, {Furlong}, {McCarthy},
  {Schaller}, \& {Theuns}}]{2015MNRAS.453..721V}
{Velliscig}, M., {Cacciato}, M., {Schaye}, J., {et~al.} 2015, \mnras, 453, 721,
  \dodoi{10.1093/mnras/stv1690}

\bibitem[{{Wang} {et~al.}(2021){Wang}, {Libeskind}, {Tempel}, {Kang}, \&
  {Guo}}]{2021NatAs...5..839W}
{Wang}, P., {Libeskind}, N.~I., {Tempel}, E., {Kang}, X., \& {Guo}, Q. 2021,
  Nature Astronomy, 5, 839, \dodoi{10.1038/s41550-021-01380-6}

\bibitem[{{Weinberg} {et~al.}(2004){Weinberg}, {Dav{\'e}}, {Katz}, \&
  {Hernquist}}]{2004ApJ...601....1W}
{Weinberg}, D.~H., {Dav{\'e}}, R., {Katz}, N., \& {Hernquist}, L. 2004, \apj,
  601, 1, \dodoi{10.1086/380481}

\bibitem[{{White}(1984)}]{1984ApJ...286...38W}
{White}, S.~D.~M. 1984, \apj, 286, 38, \dodoi{10.1086/162573}

\bibitem[{{Xia} {et~al.}(2021){Xia}, {Neyrinck}, {Cai}, \&
  {Arag{\'o}n-Calvo}}]{2021MNRAS.506.1059X}
{Xia}, Q., {Neyrinck}, M.~C., {Cai}, Y.-C., \& {Arag{\'o}n-Calvo}, M.~A. 2021,
  \mnras, 506, 1059, \dodoi{10.1093/mnras/stab1713}

\bibitem[{{York} {et~al.}(2000){York}, {Adelman}, {Anderson}, {Anderson},
  {Annis}, {Bahcall}, {Bakken}, {Barkhouser}, {Bastian}, {Berman}, {Boroski},
  {Bracker}, {Briegel}, {Briggs}, {Brinkmann}, {Brunner}, {Burles}, {Carey},
  {Carr}, {Castander}, {Chen}, {Colestock}, {Connolly}, {Crocker}, {Csabai},
  {Czarapata}, {Davis}, {Doi}, {Dombeck}, {Eisenstein}, {Ellman}, {Elms},
  {Evans}, {Fan}, {Federwitz}, {Fiscelli}, {Friedman}, {Frieman}, {Fukugita},
  {Gillespie}, {Gunn}, {Gurbani}, {de Haas}, {Haldeman}, {Harris}, {Hayes},
  {Heckman}, {Hennessy}, {Hindsley}, {Holm}, {Holmgren}, {Huang}, {Hull},
  {Husby}, {Ichikawa}, {Ichikawa}, {Ivezi{\'c}}, {Kent}, {Kim}, {Kinney},
  {Klaene}, {Kleinman}, {Kleinman}, {Knapp}, {Korienek}, {Kron}, {Kunszt},
  {Lamb}, {Lee}, {Leger}, {Limmongkol}, {Lindenmeyer}, {Long}, {Loomis},
  {Loveday}, {Lucinio}, {Lupton}, {MacKinnon}, {Mannery}, {Mantsch}, {Margon},
  {McGehee}, {McKay}, {Meiksin}, {Merelli}, {Monet}, {Munn}, {Narayanan},
  {Nash}, {Neilsen}, {Neswold}, {Newberg}, {Nichol}, {Nicinski}, {Nonino},
  {Okada}, {Okamura}, {Ostriker}, {Owen}, {Pauls}, {Peoples}, {Peterson},
  {Petravick}, {Pier}, {Pope}, {Pordes}, {Prosapio}, {Rechenmacher}, {Quinn},
  {Richards}, {Richmond}, {Rivetta}, {Rockosi}, {Ruthmansdorfer}, {Sandford},
  {Schlegel}, {Schneider}, {Sekiguchi}, {Sergey}, {Shimasaku}, {Siegmund},
  {Smee}, {Smith}, {Snedden}, {Stone}, {Stoughton}, {Strauss}, {Stubbs},
  {SubbaRao}, {Szalay}, {Szapudi}, {Szokoly}, {Thakar}, {Tremonti}, {Tucker},
  {Uomoto}, {Vanden Berk}, {Vogeley}, {Waddell}, {Wang}, {Watanabe},
  {Weinberg}, {Yanny}, {Yasuda}, \& {SDSS Collaboration}}]{2000AJ....120.1579Y}
{York}, D.~G., {Adelman}, J., {Anderson}, John~E., J., {et~al.} 2000, \aj, 120,
  1579, \dodoi{10.1086/301513}

\bibitem[{{Zel'dovich}(1970)}]{1970A&A.....5...84Z}
{Zel'dovich}, Y.~B. 1970, \aap, 5, 84

\end{thebibliography}
\bibliographystyle{aasjournal}



\end{document}